\documentclass{emulateapj}
\usepackage{graphics}

\begin{document}

\title{Supernova Simulations with Boltzmann Neutrino Transport: A Comparison
of Methods}

\author{M. Liebend\"orfer\altaffilmark{1,2,3}, M. Rampp\altaffilmark{4}, H.-Th. Janka\altaffilmark{4}, A. Mezzacappa\altaffilmark{2}} \altaffiltext{1}{CITA, University of Toronto, Toronto, Ontario M5S 3H8, Canada } \altaffiltext{2}{Physics Division, Oak Ridge National Laboratory, Oak Ridge, Tennessee 37831-6354} \altaffiltext{3}{Department of Physics and Astronomy, University of Tennessee, Knoxville, Tennessee 37996-1200} \altaffiltext{4}{Max-Planck-Institut f\"ur Astrophysik, Karl-Schwarzschild-Strasse 1, D-85741 Garching, Germany}

\begin{abstract}
Accurate neutrino transport has been built into spherically symmetric
simulations of stellar core collapse and postbounce evolution. The
results of such simulations agree that spherically symmetric models
with standard microphysical input fail to explode by the delayed,
neutrino-driven mechanism. Independent groups implemented fundamentally
different numerical methods to tackle the Boltzmann neutrino transport
equation. Here we present a direct and detailed comparison of such
neutrino radiation-hydrodynamical simulations for two codes, {\sc agile-boltztran}
of the Oak Ridge-Basel group and {\sc vertex} of the Garching group.
The former solves the Boltzmann equation directly by an implicit,
general relativistic discrete angle method on the adaptive grid of
a conservative implicit hydrodynamics code with second-order TVD advection.
In contrast, the latter couples a variable Eddington factor technique
with an explicit, moving-grid, conservative high-order Riemann solver
with important relativistic effects treated by an effective gravitational
potential. The presented study is meant to test both neutrino radiation-hydrodynamics
implementations and to provide a data basis for comparisons and verifications
of supernova codes to be developed in the future. Results are discussed
for simulations of the core collapse and post-bounce evolution of
a \( 13 \) M\( _{\odot } \) star with Newtonian gravity and a \( 15 \)
M\( _{\odot } \) star with relativistic gravity.
\end{abstract}

\section{Introduction}

Computer simulations are becoming more and more popular. They allow
investigations of physics on office desks rather than explorations
through hands-on experiments (This does not imply a transition from
hard work to gaming). In the industrial context, the two approaches
are not separable: the computer codes have to be validated. After
a computer design has been completed, its relation to reality will
inevitably be assessed in the manufacture and evaluation of prototypes.
How about the growing importance of computer simulations in astrophysics
- where are the measurements found to test aspects of a complex computer
code in idealized setups, and where are the prototypes that validate
the quality of the results in the targeted application? The first
step of code development is accompanied by the verification of partial
aspects of the code in simplified test problems where the solution
is analytically known. The code can be improved step by step. Additionally,
laboratory experiments may be used to further verify the code with
accurate measurements in idealized setups. The transition to code
validation is made when its capability to handle complicated coupled
processes is tested and the completeness of the physical description
is evaluated. In the industrial context, this is achieved by more
comprehensive experiments and measurements, or, ultimately, by the
comparison of computer designs with the properties of manufactured
prototypes. We would now be tempted to relate the validation of computer
generated results with real life prototypes in manufacturing to the
comparison of astrophysical simulations with astronomical observations.
This would, however, circumvent the goal of astrophysical simulations:
One does not assume unknown physics in industrial design. Perfect
agreement between a computer simulation and the behavior of a prototype
can indeed be seen as proof of the quality of the computer code. The
situation is different in astrophysics, where the understanding of
the physics of an event is rather the goal than the ingredient. The
comparison between simulation and observation is essential to demonstrate
the physical understanding of the event, it cannot at the same time
be used to qualify code performance. The gap in code validation between
detached analytical test calculations and the astrophysical application
can be bridged by code comparisons \citep{Calder_et_al_02}, based
on the assumption that different numerical approaches are likely to
show different strengths and weaknesses in simulations of complex
physical systems. Differences in the simulation results are an indicator
for uncertainties in the numerical methods.

In the present paper we document the detailed comparison of results
from different supernova codes. Both of our codes aim to provide a
solution to the Boltzmann neutrino transport equation in spherical
symmetry. This is achieved by fundamentally different numerical methods:
The code of the Oak Ridge-Basel group ({\sc agile-boltztran}) consists
of a general relativistic time-implicit discrete-angle (S\( _{N} \))
Boltzmann solver, which is coupled in an operator split fashion to
a general relativistic time-implicit hydrodynamics solver with a dynamical
adaptive grid. It implements a direct finite difference representation
of the Boltzmann equation \citep{Mezzacappa_Bruenn_93a, Mezzacappa_Messer_99, Liebendoerfer_Rosswog_Thielemann_02, Liebendoerfer_et_al_04}.
The Garching code, {\sc vertex}, is a one-dimensional version of a
program that was developed to perform multi-dimensional supernova
simulations with accurate ray-by-ray neutrino transport. It is based
on the explicit, moving-grid, finite-volume hydrodynamics code {\sc prometheus},
which employs a Riemann solver for constructing the solution of the
hydrodynamics equations. The neutrino transport is handled in an operator-split
step and is calculated by a variable Eddington factor closure of neutrino
energy, number, and momentum equations, where the variable Eddington
factor is derived from the formal solution of a spherically averaged
model Boltzmann equation \citep{Rampp_Janka_02}. In spherical symmetry,
there is only one {}``ray'' for the solution of the moments equations
and no spherical averaging is necessary for the model Boltzmann equation.
Therefore, a convergence in the iterations between the moments equations
and the closure from the model Boltzmann equation provides in spherical
symmetry a solution of the complete Boltzmann equation.

This work has two main goals. (i) The direct comparison of two codes
applied to the same challenging astrophysical scenario with concerted
physics (spherical symmetry, progenitor models, nuclear and weak interaction
physics, general relativistic effects). (ii) The production of reference
results to test future supernova codes in the spherical limit. Machine-readable
data files are included in the electronic edition of the Journal.

Neutrinos play a crucial role in collapsing cores of massive stars.
The loss of electron lepton number by the production and escape of
electron neutrinos determines the collapse dynamics and the position
where the supernova shock forms. Energy and lepton number transport
by neutrino diffusion also govern the evolution of the nascent neutron
star. The energy transfer by neutrinos to the medium that surrounds
the protoneutron star may revive the stalled accretion front and thus
drive a delayed explosion \citep{Wilson_85,Bethe_Wilson_85}. Neutrino
interactions moreover set the proton-to-nucleon ratio and therefore
the conditions for nucleosynthesis in the innermost supernova ejecta.
Sustained energy deposition near the protoneutron star surface causes
a post-explosion outflow of baryonic matter, the so-called neutrino-driven
wind, which is discussed as a potential site for the formation of
r-process elements \citep{Woosley_et_al_94, Takahashi_Witti_Janka_94, Sumiyoshi_et_al_00, Wanajo_et_al_01, Thompson_Burrows_Meyer_01}.

Deep inside the protoneutron star the absorption and scattering mean
free paths of neutrinos are very small and therefore neutrinos diffuse
and are in chemical equilibrium with the stellar plasma. With decreasing
density the neutrino interaction lengths become larger, before, finally
the neutrinos can stream freely. Since the reaction cross sections
rise steeply with the neutrino energy, low-energy neutrinos decouple
from the stellar background at higher densities. Most electron flavor
neutrinos emerge from the accreting material at the base of the cooling
region under semi-transparent conditions and propagate to the heating
region where their angular distribution influences the energy deposition
behind the accretion front. Neither diffusion nor free streaming is
a good approximation in this important region where neutrinos strongly
couple the dynamics of different layers on a short propagation time
scale. 

An accurate treatment of the neutrino transport and of neutrino-matter
interactions therefore requires the combination of neutrino sources
at one location with neutrino opacities at other locations as described by the
energy- and angle-dependent Boltzmann transport equation. The solution
of the Boltzmann equation is also desirable to test approximations,
the most elaborate of which are certainly multi-group flux-limited
diffusion \citep{Bruenn_85, Myra_et_al_87, Bruenn_DeNisco_Mezzacappa_01}
and two-moment closure schemes \citep{Bludman_Cernohorsky_95, Smit_Cernohorsky_Dullemond_97}.
With the growing computer capability it has become feasible to provide
solutions to the Boltzmann transport equation not only for the collapse
phase \citep{Mezzacappa_Bruenn_93a}, but also in consistent dynamical
simulations of the post-bounce evolution \citep{Rampp_Janka_00, Mezzacappa_et_al_01, Liebendoerfer_et_al_01, Thompson_Burrows_Pinto_03}.

The paper is organized as follows. We will describe in Sect.~\ref{section_model_description}
the stellar models and the physical ingredients that constitute the
problem to be solved by our numerical methods. In Sect.~\ref{section_technical}
we will briefly resume characteristic features and capabilities of
both neutrino radiation-hydrodynamics codes. In Sect.~\ref{section_results}
our results for the two considered stellar models will be discussed
with special focus on the differences between the runs. In Sect.~\ref{section_summary}
we shall summarize our findings and draw conclusions.

\section{Description of the Model}

\label{section_model_description}

\subsection{Progenitors}

We present in this paper two different models. One of them includes
only the most essential physical ingredients; e.g., the transport
of electron flavor neutrinos and antineutrinos, but not the heavy-lepton
neutrinos and antineutrinos. It is meant to serve as a guideline for
future code development and transport approximations. The model is
based on a \( 13 \) M\( _{\odot } \) progenitor of \citet{Nomoto_Hashimoto_88}.
The \( 13 \) M\( _{\odot } \) progenitor model has a tradition in
supernova simulations, its exceptionally small iron core sustained
the hope to produce prompt explosions. We call this run N13, as it
is based on Newtonian gravity. A second run was launched from a \( 15 \)
M\( _{\odot } \) progenitor of \citet{Woosley_Weaver_95}. This progenitor
has been widely used in supernova investigations, as it provides a
model of a massive star in the middle of the mass range that is expected
to end its life in a supernova. The run takes into account all neutrino
flavors with {}``standard'' input physics as listed in \citet{Bruenn_85}.
General relativistic effects are included in this physically more
complete run, G15, and the input physics has been extended by ion-ion
correlations and nucleon-nucleon bremsstrahlung.

\subsection{Radiation Hydrodynamics in Spherical Symmetry}

The stellar progenitor model is evolved in time by means of the hydrodynamics
and neutrino transport equations. Since there is no danger of grid
entangling in spherical symmetry, we make use of the freedom in {\sc agile-boltztran}
to choose orthogonal comoving space-time coordinates and describe
the interactions in the collision term in the most convenient comoving
frame for the neutrino four-momentum (see e.g., \citet{Cardall_Mezzacappa_03}
for a generalized discussion of coordinate choices for the radiation
transport). In the metric of \citet{Misner_Sharp_64, May_White_66},
\begin{equation}
\label{eq_comoving_metric}
ds^{2}=-\alpha ^{2}\left( cdt\right) ^{2}+\left( \frac{1}{\Gamma }\frac{\partial r}{\partial a}\right) ^{2}da^{2}+r^{2}\left( d\vartheta ^{2}+\sin ^{2}\vartheta d\varphi ^{2}\right) ,
\end{equation}
 the equations of hydrodynamics in spherical symmetry in the presence
of a radiation field can be written in the form \citep{Lindquist_66, Liebendoerfer_Mezzacappa_Thielemann_01}

\begin{eqnarray}
\lefteqn{\frac{\partial }{\partial t}\left[ \frac{\Gamma }{\rho }\right] = \frac{\partial }{\partial a}\left[ 4\pi r^{2}\alpha u\right]} & \qquad & \label{eq_density} \\
\lefteqn{\frac{\partial }{\partial t}\left[ \Gamma \left( 1+\frac{e+J}{c^{2}}\right) +\frac{uH}{c^{4}}\right] = } \nonumber \\
 &   & -\frac{\partial }{c^{2}\partial a}\left[ 4\pi r^{2}\alpha \left( up+u\rho K+\Gamma \rho H\right) \right] \label{eq_energy} \\
\lefteqn{\frac{\partial }{\partial t}\left[ u\left( 1+\frac{e+J}{c^{2}}\right) +\frac{\Gamma H}{c^{2}}\right] = } \nonumber \\
 &   & -\frac{\partial }{\partial a}\left[ 4\pi r^{2}\alpha \left( \Gamma p+\Gamma \rho K+\frac{u\rho H}{c^{2}}\right) \right] \nonumber \\
 &   & {} - \frac{\alpha }{r}\left[ \left( 1+\frac{e+3p/\rho +J+3K}{c^{2}}\right) \frac{Gm}{r}\right. \nonumber \\
 &   & {} - \left. \left( 1-\frac{2Gm}{rc^{2}}\right) \left( J-3K\right) -2\left( \frac{p}{\rho }+K\right) \right. \nonumber \\
 &   & {} + \left. 8\pi r^{2}\frac{G}{c^{2}}\left( \left( 1+\frac{e+J}{c^{2}}\right) \left( p+\rho K\right) -\frac{\rho H^{2}}{c^{4}}\right) \right] \label{eq_momentum} \\
\lefteqn{\frac{\partial }{\partial a}\left[ \frac{4\pi }{3}r^{3}\right]  = \frac{\Gamma }{\rho }\label{eq_volume}} \\
\lefteqn{\frac{\partial m}{\partial a} = \Gamma \left( 1+\frac{e+J}{c^{2}}\right) +\frac{uH}{c^{4}}\label{eq_mass} } \\
\lefteqn{\frac{\partial }{\partial t}\left[ \frac{1}{4\pi r^{2}\rho }\frac{H}{c^{4}}\right] = } \nonumber \\
 &   & -\left( 1+\frac{e+J}{c^{2}}\right) \frac{\partial \alpha }{\partial a}-\frac{1}{\rho c^{2}}\frac{\partial }{\partial a}\left[ \alpha \left( p+\rho K\right) \right] \nonumber \\
 &   & {} + \frac{\alpha \Gamma }{4\pi r^{3}\rho c^{2}}\left( J-3K\right) . 
\label{eq_lapse}\end{eqnarray}

The metric is based on an enclosed baryon mass label, \( a \), and
a coordinate time, \( t \). It refers to the areal radius, \( r \),
the {}``Lorentz'' factor, \( \Gamma =\sqrt{1+(u/c)^{2}-2Gm/(rc^{2})} \),
and the lapse function, \( \alpha  \). The angles \( \vartheta  \)
and \( \varphi  \) describe a two-sphere. The quantity \( u=\alpha ^{-1}\partial r/\partial t \)
takes the place of a fluid velocity and \( m \) is the enclosed gravitational
mass, proportional to the enclosed total energy. Both of our codes
split these equations into hydrodynamics equations and transport equations.
{\sc vertex} employs Eulerian coordinates which can be obtained by
a coordinate transformation. The fluid is specified in its rest frame
by the rest mass density, \( \rho  \), specific internal energy,
\( e \), and electron number fraction, \( Y_{e} \), for which an
additional evolution equation (lepton number conservation) is solved.
The hydrodynamics equations are closed by the equation of state, which
gives the pressure, \( p \), as a function of density, internal energy,
and electron number. The zeroth, \( J \), first, \( H/c \), and
second, \( K \), angular moment of the monochromatic neutrino intensity
(normalized by the rest-mass density) are determined by the transport
equation which includes the interactions listed in Table \ref{tab:reactions_n13}
in the collision integral. Eq. (\ref{eq_lapse}) determines the lapse
function. Eq. (\ref{eq_mass}) allows us to integrate outward from
\( a=0 \) to obtain the total energy; it translates to the Poisson
equation in the Newtonian limit. Eq. (\ref{eq_volume}) defines the
general relativistic analogue to the Newtonian volume. Eq. (\ref{eq_momentum})
describes the change of radial momentum; to leading order it is proportional
to \( -Gm/(cr)^{2}+(J-3K)/r \). Eq. (\ref{eq_energy}) describes
the evolution of the total energy. Note that there are no contributions
from terms of zeroth and first order in \( c \). Eq. (\ref{eq_density})
relates the evolution of the specific volume to the divergence of
the velocity field.

\subsection{Nuclear and Weak Interaction Physics Input}

The equation of state describes the thermodynamic state of a fluid
element based on density, \( \rho  \), temperature, \( T \), and
the composition. The relation between the specific internal energy
and the pressure closes the system of hydrodynamics equations. For
this comparison we use the equation of state of \citet{Lattimer_Swesty_91}.
It assumes nuclear statistical equilibrium and is based on a liquid
drop model for a representative nucleus with atomic number \( A \)
and charge \( Z \), surrounded by free alpha particles, protons,
and neutrons. These baryons are immersed in an electron and positron
gas that equilibrates with a photon gas by the pair creation process.
The incompressibility modulus can be adjusted. We use a value of \( K=180 \)
MeV. Above nuclear density, where no isolated individual nuclei are
present, the transition to a proton-neutron-electron gas is made by
a Maxwell phase transition. In any of these cases, the nuclear composition
at given temperature and density is determined by the specification
of the electron fraction \( Y_{e} \). At low densities/temperatures
the {\sc vertex} code switches to an equation of state that considers
electrons, positrons, photons, nucleons and nuclei consistent with
the composition used in the progenitor model \citep{Rampp_Janka_02}.
The switch is triggered in N13 and G15 by a density threshold of \( \rho <6\times 10^{7}\,  \)g/cm\( ^{3} \).
{\sc agile-boltztran} applies the same switch in the run N13, but
considers in the low density domain only one nucleus, \( ^{28} \)Si.
In run G15, silicon is converted to NSE under energy conservation
at a burning temperature of \( 0.44 \) MeV \citep{Mezzacappa_et_al_01}.

As for the neutrino-matter interactions, we have chosen to use the
conventional ({}``standard'') opacities, i.e., a description which
follows closely the one detailed by \citet{Bruenn_85, Mezzacappa_Bruenn_93c}.
Note, however, that there are small differences in the neutrino description
employed by the two groups. While {\sc agile-boltztran} treats the
\( \mu /\tau  \) neutrinos and antineutrinos separately, they are
combined to one species in {\sc vertex}. In order to save computer
time, usually only electron neutrinos are considered in the {\sc vertex}
calculations during the core collapse phase. Tests have shown that
taking into account also electron antineutrinos and the heavy-lepton
neutrinos leads to only minuscule differences in this phase of the
supernova evolution (their inclusion in the postbounce phase, however,
is important as demonstrated by the comparison between models N13
and G15 in Sect. \ref{section_results}). The Garching group routinely
includes nucleon-nucleon bremsstrahlung as a source (or sink) for
neutrino-antineutrino pairs. The decrease of coherent neutrino scattering
off heavy ions due to ion-ion correlations and electron screening
\citep{Itoh_75,Horowitz_97,Bruenn_Mezzacappa_97} is also taken into
account. These improvements have been switched off in the N13 models
and added to {\sc agile-boltztran} such that they are consistently
included in both G15 runs. In both codes, the implementation of the
ion-ion correlations has been updated with the structure function
given in \citet{Itoh_et_al_04}. Because of the large mass contrast
between species with \( A\leq 4 \) and the representative heavy nucleus
we omit the somewhat arbitrary averaging of the effect over species
and consider only the representative nucleus for the calculation of
the ion separation.
\begin{table}
\caption{Overview of all neutrino-matter interactions considered in the N13
runs.\label{tab:reactions_n13}}
\centerline{
\begin{tabular}{rcll}
\hline 
\multicolumn{3}{l}{
Reaction\protect\footnote{In the first column the symbol \protect\( \nu \protect \) represents
any of the neutrinos \protect\( \nu _{\mathrm{e}}\protect \), \protect\( \bar{\nu }_{\mathrm{e}}\protect \)
(heavy-lepton neutrinos are neglected in N13). The symbols \protect\( \mathrm{e}^{-}\protect \),
\protect\( \mathrm{e}^{+}\protect \), \protect\( \mathrm{n}\protect \),
\protect\( \mathrm{p}\protect \) and \protect\( A\protect \) denote
electrons, positrons, free neutrons and protons, and heavy nuclei,
respectively, the symbol \protect\( \mathrm{N}\protect \) means \protect\( \mathrm{n}\protect \)
or \protect\( \mathrm{p}\protect \). The references point to papers
where information can be found about the approximations employed in
the rate calculations. Details about the numerical implementation
can be found in the methodical papers of \citet{Rampp_Janka_02} and
\citet{Mezzacappa_Messer_99, Liebendoerfer_et_al_04}, respectively.
The G15 runs additionally include reactions with \protect\( \mu \protect \)-
and \protect\( \tau \protect \)-neutrinos and their antiparticles,
ion-ion correlation effects in neutrino-nuclei interactions, and nucleon-nucleon
bremsstrahlung.}}
&
 Reference \\
\hline
\( \nu A \)&
\( \rightleftharpoons  \)&
 \( \nu A \)&
\citet{Bruenn_85} (no ion-ion correlations!)\\
\( \nu \mathrm{N} \)&
\( \rightleftharpoons  \)&
 \( \nu \mathrm{N} \)&
\citet{Bruenn_85, Mezzacappa_Bruenn_93c}\\
\( \nu _{\mathrm{e}}\mathrm{n} \)&
\( \rightleftharpoons  \)&
 \( \mathrm{e}^{-}\mathrm{p} \)&
\citet{Bruenn_85, Mezzacappa_Bruenn_93c}\\
\( \bar{\nu }_{\mathrm{e}}\mathrm{p} \)&
\( \rightleftharpoons  \)&
 \( \mathrm{e}^{+}\mathrm{n} \)&
\citet{Bruenn_85, Mezzacappa_Bruenn_93c}\\
\( \nu _{\mathrm{e}}A' \)&
\( \rightleftharpoons  \)&
 \( \mathrm{e}^{-}A \)&
\citet{Bruenn_85, Mezzacappa_Bruenn_93c}\\
\( \nu \bar{\nu } \)&
\( \rightleftharpoons  \)&
 \( \mathrm{e}^{-}\mathrm{e}^{+} \)&
\citet{Bruenn_85} \\
\( \nu \mathrm{e}^{\pm } \)&
\( \rightleftharpoons  \)&
 \( \nu \mathrm{e}^{\pm } \)&
\citet{Mezzacappa_Bruenn_93c}; \\
& & & \citet{Cernohorsky_94}\\
\hline
\end{tabular}}
\end{table}

\section{Numerical Methods}

\label{section_technical}The two codes follow very different approaches
to evaluate the radiation moments \( J \), \( H \), and \( K \).
Contrary to flux-limiting and {}``gray'' transport methods, neither
of our methods needs to make assumptions about the angular or the
spectral distribution of the radiation field. Important features and
implementation details of the two codes are summarized in the following
subsections.

\subsection{{\sc agile-boltztran}}

The concept and first implementation of {\sc boltztran} has been developed
in \citet{Mezzacappa_Bruenn_93a, Mezzacappa_Bruenn_93b, Mezzacappa_Bruenn_93c}
for core collapse simulations in the order \( v/c \) approximation.
Essential for the computational efficiency of the implicit scheme
is the storage of the interactions in a dynamical table which delivers
consistent derivatives of all cross sections and thermodynamical quantities
for the Newton-Raphson iterations in the solution of the nonlinear
equations \citep{Mezzacappa_Messer_99, Messer_00}. For the highly
dynamical situation after bounce, {\sc boltztran} has been coupled
to the hydrodynamics code {\sc agile}. The finite differencing has
been revised for energy conservation and extended to solve the general
relativistic Eqs. (\ref{eq_density}-\ref{eq_boltzmann}) \citep{Liebendoerfer_00, Liebendoerfer_Rosswog_Thielemann_02, Liebendoerfer_et_al_04}.

\subsubsection{Hydrodynamics}

The hydrodynamics part of the Lagrangian Eqs. (\ref{eq_density}-\ref{eq_lapse})
is solved by implicit conservative finite differencing. One strength
is the implementation of a dynamically moving adaptive grid following
\citet{Winkler_Norman_Mihalas_84} and \citet{Dorfi_Drury_87}. In the
general relativistic extension, it is equivalent to a resolution-dependent
choice of shift vectors, that allow a continuous coordinate translation
in the radial direction during the evolution of the model. Artificial
viscosity has been included in a consistent, but causality violating
manner based on the tensor viscosity of \citet{Tscharnuter_Winkler_79}.
It provides the mechanism for energy dissipation in the shock front
and defines the shock width such that the number of attracted adaptive
grid points does not grow beyond limits. A major advantage of the
adaptive grid is the dynamical allocation of computational zones to
regions where they are needed. The zoning for hydrodynamics and neutrino
transport is always kept congruent for consistency reasons. During
the evolution of the model, one group of grid points follows the accretion
front, while another resolves the steep density gradient between the
outer layers of the protoneutron star and the infalling matter, where
most of the electron flavor neutrinos stream away. The use of artificial
viscosity is not a disadvantage of the method, as it only plays a
minor local role in stabilizing the shock front (the shock width is
set to a few percent of the shock radius and captured by \( \sim 10 \)
moving grid points). A disadvantage of the adaptive grid approach
in an earlier implementation (see \citet{Liebendoerfer_Rosswog_Thielemann_02}
for a detailed description) is the numerical diffusion introduced
by the first-order advection in regions where the adaptive grid does
not concentrate its zones, e.g. in a rarefaction wave or at sharp
discontinuities of the composition. During this comparison, improvement
has been achieved by upgrading {\sc agile} with second order total
variation diminishing (TVD) advection based on a Van Leer flux limiter.

\subsubsection{Neutrino Transport}

The neutrino transport part, {\sc boltztran}, solves the Boltzmann
transport equation,\begin{eqnarray}
C & = & \frac{\partial F}{\alpha c\partial t}+\frac{\mu }{\alpha }\frac{\partial }{\partial a}\left[ 4\pi r^{2}\alpha \rho F\right] \nonumber \\
 & + & \Gamma \left( \frac{1}{r}-\frac{1}{\alpha }\frac{\partial \alpha }{\partial r}\right) \frac{\partial }{\partial \mu }\left[ \left( 1-\mu ^{2}\right) F\right] \nonumber \\
 & + & \left( \frac{\partial \ln \rho }{\alpha c\partial t}+\frac{3u}{rc}\right) \frac{\partial }{\partial \mu }\left[ \mu \left( 1-\mu ^{2}\right) F\right] \nonumber \\
 & - & \mu \Gamma \frac{1}{\alpha }\frac{\partial \alpha }{\partial r}\frac{1}{E^{2}}\frac{\partial }{\partial E}\left[ E^{3}F\right] \nonumber \\
 & + & \left( \mu ^{2}\left( \frac{\partial \ln \rho }{\alpha c\partial t}+\frac{3u}{rc}\right) -\frac{u}{rc}\right) \frac{1}{E^{2}}\frac{\partial }{\partial E}\left[ E^{3}F\right] ,\label{eq_boltzmann} 
\end{eqnarray}
in a finite difference representation implementing the discrete ordinates,
or S\( _{N} \), method. The evolved quantity is the neutrino distribution
function, \( F(t,a,\mu ,E) \), as a function of time \( t \), rest
mass \( a \) enclosed in a sphere at radius \( r \), propagation
angle cosine \( \mu  \) with respect to radial direction, and neutrino
energy \( E \). Neutrinos in specific beams are created and destroyed
according to the collision term, \( C \), which includes the interactions
listed in Table \ref{tab:reactions_n13}. It is assumed that the neutrinos
propagate freely between interactions. The free-particle motion along
geodesics between collisions introduces the many correction terms
apparent on the right hand side of Eq. (\ref{eq_boltzmann}). They
stem from the use of spherical coordinates in combination with a description
of the neutrino phase space in a comoving frame. Nevertheless, all
terms can be labeled with a physical effect. In order of appearance
in Eq. (\ref{eq_boltzmann}) these are, the time derivative of the
neutrino distribution function, the propagation of neutrinos, the
angle correction due to neutrino propagation, the angular aberration
correction due to observer motion, the frequency shift in the gravitational
potential, and the Doppler frequency shift due to observer motion.
It is essential for the successful finite difference representation
of Eq. (\ref{eq_boltzmann}) that it is upward compatible with simple
special cases of the transport equation. Basic physical properties
can be determined by the evaluation of expectation values with the
neutrino distribution functions for various operators. The finite
difference representation should support, e.g., the diffusion limit,
total energy conservation, and the conservation of lepton number.

\subsubsection{Parameter settings}

Both runs, N13 and G15, were performed with \( 103 \) adaptive spatial
zones ranging from the center of the progenitor star to about \( 7000 \)
km. A constant pressure boundary condition was used at the barely
moving surface. The neutrino energy was resolved with \( 20 \) geometrically
increasing energy groups, the first centered at \( 3 \) MeV and the
last at \( 300 \) MeV. The propagation angle has been discretized
with \( 6 \) angles suitable for Gaussian quadrature. Roughly \( 3000 \)
time steps have been used for collapse and \( 7000 \) for the postbounce
phase. The run N13 has been evolved with an order \( v/c \) approximation
of Eqs. (\ref{eq_density}-\ref{eq_boltzmann}) and run G15 with the
general relativistic equations.

\subsection{{\sc vertex}}

Independently from the efforts of the Oak Ridge-Basel collaboration
the Garching supernova group has treated the Boltzmann transport problem
for neutrinos in core-collapse supernovae with a new variable Eddington
factor method \citep{Rampp_00, Rampp_Janka_00, Rampp_Janka_02}, and
has coupled it to the {\sc prometheus} hydrodynamics code. The combined
program allows for spherically symmetric \citep{Rampp_Janka_00, Rampp_Janka_02}
as well as multi-dimensional simulations \citep{Janka_et_al_04, Buras_et_al_03}.
The spherically symmetric {}``core'' of the program, which was used
for the calculations described below, will be referred to by the name
{\sc vertex} (\textbf{V}ariable \textbf{E}ddington factor \textbf{R}adiative
\textbf{T}ransfer for supernova \textbf{EX}plosions).

\subsubsection{Hydrodynamics}

The integration of the equations of hydrodynamics is performed with
the Newtonian finite-volume code {\sc prometheus} \citep{Fryxell_Mueller_Arnett_89}
which was supplemented by additional problem specific features \citep{Keil_97}.
{\sc prometheus} is a direct Eulerian, time-explicit implementation
of the Piecewise Parabolic Method (PPM) of \citet{Colella_Woodward_84}.
As a Godunov scheme of third order in space and second-order in time
with a Riemann solver, it is particularly well suited for following
discontinuities in the fluid flow like shocks or boundaries between
layers of different chemical composition. A notable advantage is its
capability of tackling multi-dimensional problems with high computational
efficiency and numerical accuracy. The code makes use of the {}``Consistent
Multifluid Advection (CMA)'' method \citep{Plewa_Mueller_99} for
ensuring accurate advection of different chemical components in the
fluid, and switches from the original PPM method to the more diffusive
HLLE solver of \citet{Einfeldt_88} in the vicinity of strong shocks
to avoid spurious oscillations (the so-called {}``odd-even decoupling''
phenomenon) when such shocks are aligned with one of the coordinate
lines in multidimensional simulations \citep{Quirk_94, Kifonidis_00, Plewa_Mueller_01}.

\subsubsection{Neutrino transport}

The variable Eddington factor scheme for the neutrino transport, its
coupling to the hydrodynamics part, and application to a number of
test problems is described in much detail elsewhere \citep{Rampp_Janka_02}.
Here we will only briefly summarize the characteristic features of
the method. The coupled set of equations of hydrodynamics (Eqs. (1)--(4)
in \citet{Rampp_Janka_02}) and radiation transport (Eqs. (6)--(8)
ibidem) is equivalent to Eqs. (\ref{eq_density})-(\ref{eq_lapse})
in the order \( (v/c) \) limit. The equations are also split into
a {}``hydrodynamics part'' and a {}``neutrino part'' and solved
independently in subsequent ({}``fractional'') steps. But the hydrodynamics
and the transport solver can use radial grids and/or time steps that
are different from each other.

In the neutrino transport method the integro-differential character
of the Boltzmann equation is tamed by applying a variable Eddington
factor closure to the neutrino energy and momentum equations (and
the simultaneously integrated first and second order moments equations
for neutrino number). For this purpose the variable Eddington factor
is determined from the formal solution of the Boltzmann equation on
so-called {}``tangent rays''. They coincide with radiation characteristics
in Newtonian geometry. The system of the Boltzmann equation and its
moments equations is iterated until convergence is achieved. The integration
of the transport equations is implicit in time.

General relativistic effects are treated only approximately in the
code \citep{Rampp_Janka_02}. The current version contains a modification
of the gravitational potential by including correction terms due to
pressure and energy of the stellar medium and neutrinos, which are
deduced from a comparison of the Newtonian and relativistic equations
of motion. The neutrino transport contains gravitational redshift
and time dilation, but ignores the distinction between coordinate
radius and proper radius. This simplification is necessary for coupling
the transport code to the basically Newtonian hydrodynamics. We shall
demonstrate in this paper that these approximations work satisfactorily
well for the core collapse and the early postbounce phase (see Sect.
\ref{section_G15}). Moderate quantitative but no qualitative differences
from the full relativistic treatment are mainly found at late times
after the accretion front has started to retreat.

\subsubsection{Parameter settings}

For the neutrino transport in the N13 run an Eulerian radial grid
with 235 radial zones (255 tangent rays) spaced logarithmically between
0 and 10\,000 km was used. The neutrino spectrum between 0 and 380
MeV was discretized with 21 geometrically zoned energy bins, the center
of the first bin being located at 2 MeV. The hydrodynamics, on the
other hand, was solved on a radial grid of 400 zones which are moved
with the stellar fluid during core collapse. Shortly after core bounce
both radial grids were rezoned such that inside of a radius of 400
km the zones of the transport grid coincide with those of the hydro
grid. For the post-bounce evolution the coordinates of the hydrodynamics
grid (as well as those of the transport grid) remained fixed in time.
Concerning the resolution and definition of numerical grids the same
parameters were chosen in the G15 run with the exception that 19 energy
bins between 0 and 380 MeV were used instead of 21 groups and a rezoning
of the radial grid was necessary at \( \sim 200 \) ms after bounce.
The new grid (hydrodynamics and neutrino transport) employs 40 additional
radial zones in order to adequately represent the steepening density
gradients at the surface of the nascent neutron star.

\section{Comparison of the Results}

\label{section_results}The simulations produce a sizeable amount
of data, even if they are confined to one spatial dimension. Hence,
we start with an overview of what we are going to compare and how
the comparable quantities are derived from the code-specific results.
During collapse and bounce it is quite natural to choose the enclosed
baryon mass as spatial coordinate. It labels individual fluid elements
and allows one to trace the history of each fluid element. Some tenths
of milliseconds after bounce, the neutron star becomes rather static
with mass accretion being essentially stationary. Therefore the presentation
of the quantities as functions of radius is more appropriate. We present
the three independent quantities that determine the thermodynamic
state of the fluid element as measured by an observer comoving with
the fluid: the rest mass density, the entropy per baryon, and the
electron fraction. Furthermore, the radial velocity of the fluid element
is displayed.

Note that in case of the relativistic model, G15, the coordinate independent
change in areal radius per proper time, \( u=\alpha ^{-1}\partial r/\partial t \),
is plotted in the (b)-panels of Figs.~\ref{fig_B15_0}, \ref{fig_B15_1},
\ref{fig_B15_2}, and \ref{fig_B15_3} for the \textsc{agile-boltztran}
run, whereas the velocity, \( v=\partial r/\partial t \), is unchanged
from the Newtonian case in the general relativistic approximation
of \textsc{vertex}. Typical deviations of the metric coefficients
from unity are shown in Table \ref{tab_metric}.
\begin{table}
\caption{Metric coefficients\label{tab_metric}}
\centerline{
\begin{tabular}{ccccccc}
\hline 
\( t_{pb} \) {[}ms{]}&
 \( \alpha _{\mbox {max}} \)&
 \( \alpha _{\nu } \)&
 \( \alpha _{s} \)&
 \( \Gamma _{\mbox {max}} \)&
 \( \Gamma _{\nu } \)&
 \( \Gamma _{s} \)
\protect\footnote{Listed are the lapse function, \protect\( \alpha \protect \), and
the metric coefficient, \protect\( \Gamma \protect \), by their deviations
from unity in percent at several post-bounce times \protect\( t_{pb}\protect \)
during the G15 simulation with {\sc agile-boltztran}. The indices
of \protect\( \alpha \protect \) and \protect\( \Gamma \protect \)
refer to chosen locations where the values are given. The index \protect\( s\protect \)
corresponds to the position outside of the shock or accretion front,
the index \protect\( \nu \protect \) corresponds to the neutrinosphere
of the heavy lepton neutrinos, and the index max stands for the maximum
value in the whole star.}
\\
\hline
-1&
 4&
 1&
 -&
 2&
 1&
 -\\
 0&
 15&
 1&
 3&
 8&
 2&
 5\\
 1&
 15&
 1&
 1&
 7&
 1&
 2\\
 5&
 15&
 1&
 1&
 7&
 1&
 1\\
 10&
 15&
 2&
 1&
 7&
 2&
 1\\
 50&
 17&
 3&
 0&
 7&
 3&
 1\\
 100&
 18&
 3&
 0&
 8&
 3&
 0\\
 250&
 23&
 6&
 0&
 10&
 6&
 1 \\
\hline
\end{tabular}}
\end{table}
In summary, these deviations are of order \( 1\% \) in the preshock
region, increase from \( 1\% \) to \( 6\% \) at the neutrinosphere
during the simulation, and reach maximum values around \( 20\% \)
at the center of the star in case of the lapse function and around
\( 10\% \) close to the center in the case of \( \Gamma  \).

The neutrino transport quantities are represented by the neutrino
luminosity and rms energy profiles as measured in the fluid rest frame.
We also discuss selected quantities as functions of time, e.g. the
trajectories of fluid elements, the position of the (accretion-)shock,
the conditions at the center, or the neutrino luminosities and rms
energies sampled at \( 500 \) km radius (as an approximation to radial
infinity). The physical time in both runs is synchronized at bounce
\( (t=0) \), the moment when the central density reaches a local
maximum immediately before the shock is formed. Negative times in
the simulations therefore point to instances before bounce. We define
the shock position, \( R_{s} \), by the maximum of the velocity divergence
(i.e. maximum compression, cf. Eq. (\ref{eq_density})),\[
x(R_{s})=\max _{r}\left[ x(r)\right] ,\quad x=-\frac{\partial (4\pi r^{2}u)}{\partial V}.\]
 The luminosities and rms energies are given in the comoving frame
as a function of the neutrino phase-space distribution function \( F(t,a,\mu ,E) \)
according to\begin{eqnarray*}
L(t,a) & = & 4\pi r^{2}\rho \frac{2\pi c}{(hc)^{3}}\int F(t,a,\mu ,E)E^{3}dE\mu d\mu \\
\langle \epsilon (t,a)\rangle _{rms} & = & \sqrt{\frac{\int F(t,a,\mu ,E)E^{4}dEd\mu}{\int F(t,a,\mu ,E)E^{2}dEd\mu} }.
\end{eqnarray*}
 In the general relativistic case, the definition of \( F(t,a,\mu ,E) \)
in Eq. (\ref{eq_boltzmann}) implies that the luminosity at radius
\( r \) must be interpreted as originating from a neutrino number
per proper time crossing a mass shell as measured by an observer comoving
with the shell. The neutrinos carry energies, \( E \), which are
also measured in the comoving frame.

We compare time slices in our runs at three crucial instances in the
postbounce evolution: Bounce (\( t_{pb}=0 \) ms), burst (\( t_{pb}=3 \)
ms), and broil (\( t_{pb}>100 \) ms). The importance of core bounce,
the instance when the infall is reversed at the center due to the
strong repulsive forces above nuclear density, needs no further explanation.
The time slice at \( 3 \) ms not only captures the launch of the
electron neutrino burst, but also the interesting phase when the dynamical
bounce-shock stalls (i.e., the postshock velocities become negative).
At this time, long before any neutrino heating can take place, the
infalling material does no longer change the direction of its velocity
at the shock front. After deceleration at the accretion shock it continues
to drift toward the center of the star. Puffed-up by the dissipation
of the kinetic energy acquired during infall, however, the net volume
of accumulated shock-heated material still increases such that the
accretion front continues to expand to larger radii in a quasi-stationary
manner. Only after the accretion front has reached a position farther
out in the gravitational well, where less kinetic energy is dissipated
in the shock, do the temperature difference between the receding hot
neutrinospheres and the cooler postshock matter become favorable for
neutrino heating. These conditions are not met before a time of \( 50 \)
ms after bounce. In spherically symmetric simulations the accretion
front reaches a maximum radius around \( 150 \) km at about \( 100 \)
ms after bounce and recedes slowly thereafter. Therefore, we choose
a third snapshot in our comparison at this late phase where neutrino
cooling \emph{and} heating influence the quasi-stationary evolution.

\subsection{Hydrodynamics}

As described above, the dynamical simulations are based on the two
very different hydrodynamics codes \textsc{agile} and \textsc{prometheus}.
In order to disentangle hydrodynamics differences from neutrino transport
differences in our results it is helpful to perform a comparison of
an adiabatic collapse where all neutrino interactions are suppressed.
We found a simple test case that poses similar challenges to the hydrodynamics
algorithms as the case with full transport. We take the progenitor
model of run N13 and replace the electron fraction and entropy as
a function of enclosed mass by the values obtained at bounce in the
N13 run with full transport. By this measure, the Chandrasekhar mass
at core bounce, which depends on electron fraction and temperature,
is imposed already at the beginning of the simulation. As expected,
the adiabatic collapse of this modified progenitor leads to bounce
and shock formation at a similar mass coordinate as in the simulation
with full transport, and therefore to similar conditions around bounce.

\begin{figure}
{\centering \resizebox*{0.5\textwidth}{!}{\includegraphics{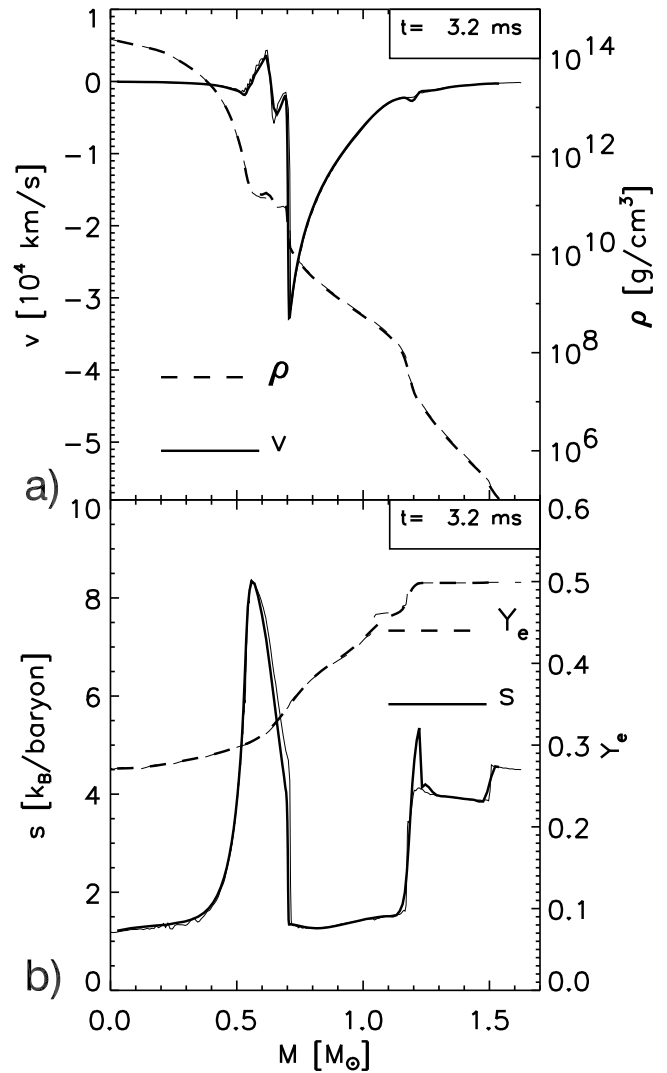}} \par}

\caption{Snapshots at \protect\( 3\protect \) ms after bounce for the adiabatic
collapse of the modified N13 progenitor. Data from the \textsc{agile}
simulation are drawn with thick lines. Data from the \textsc{prometheus}
simulation are drawn with thin lines. Panel (a) shows the velocity
(solid lines) and density (dashed lines) profiles, panel (b) the entropy
(solid lines) and electron fraction (dashed lines). The hydrodynamics
simulations agree well. The shock strength decays slightly faster
in \textsc{agile} than in \textsc{prometheus}. The weaker shock in
\textsc{agile} tends to propagate more slowly in mass and to produce
a smaller postshock entropy.\label{fig_H13}}
\end{figure}

Figure \ref{fig_H13} 
shows the situation at \( 3 \) ms after bounce. This is the critical
time when the dynamical shock in the full models stalls to turn into
an accretion front. We find very good agreement in the density and
velocity profiles. The latter show the same timing and amplitude of
reflected sound waves in the ringing neutron star. Also the entropy
profiles agree well. The entropy profile contains information about
the evolution of the shock strength because the otherwise conserved
entropy of a fluid element can only change due to the dissipation
of kinetic energy when matter passes through the shock front. The
profile shows that both the shock in \textsc{agile} and the shock
in \textsc{prometheus} start with a similar strength (almost identical
entropy peak at \( 0.55 \) M\( _{\odot } \)). In the further evolution,
however, the shock in \textsc{agile} decays slightly faster, producing
lower postshock entropies. The weaker shock propagates more slowly
with respect to the mass coordinate so that a small offset of the
shock position becomes visible at \( 3 \) ms after bounce.

\subsection{Newtonian 13 M\protect\( _{\odot }\protect \) Model}

\begin{figure*}
{\centering \resizebox*{1\textwidth}{!}{\includegraphics{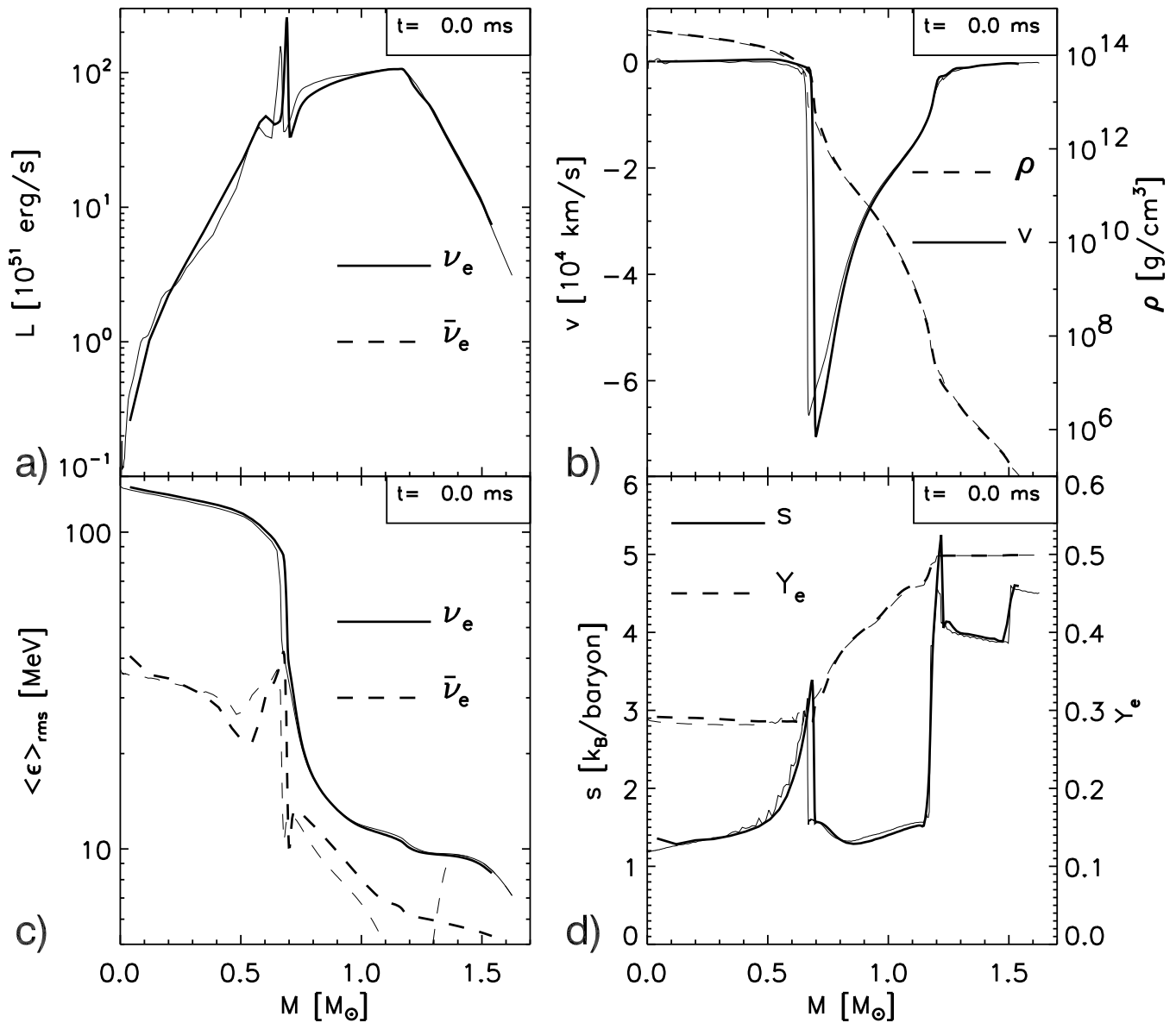}} \par}

\caption{Snapshots at bounce for model N13. Data from the \textsc{agile-boltztran}
simulation are drawn with thick lines. Data from the \textsc{vertex}
simulation are drawn with thin lines. Panel (b) shows the velocity
(solid lines) and density (dashed lines) profiles, panel (d) the entropy
(solid lines) and electron fraction (dashed lines). The neutrino luminosities
are given in panel (a) and the neutrino rms energies in panel (c).
Solid lines refer to electron neutrinos and dashed lines to electron
antineutrinos. The central electron fraction in \textsc{vertex} is
smaller than in \textsc{agile-boltztran} and the shock forms at a
slightly smaller enclosed mass. The high electron degeneracy during
collapse suppresses electron antineutrino production so that the corresponding
luminosity at bounce is below the threshold of panel (a).\label{fig_B13_0}}
\end{figure*}

For the runs that include neutrino transport, we start the comparison
with an investigation of differences in the N13 model, in which only
\( \nu _{e} \) and \( \bar{\nu }_{e} \) are taken into account.
Figure \ref{fig_B13_0}
presents the conditions at bounce. The neutrino luminosities are shown
in panel (a). We find transient differences of order \( 30 \)\% in
the electron neutrino luminosities at bounce in the diffusive regime
inside of the nascent shock. The luminosities are in good agreement
outside of the shock front and the electron neutrino rms energies
in panel (c) agree well in all domains. The electron antineutrino
rms energies are very uncertain at this time because the antineutrino
abundance is negligible under the electron-degenerate conditions at
bounce. Important quantities at bounce are the entropy and the electron
fraction in panel (d) because they determine the size of the causally
connected homologous core. A shock forms when the outgoing pressure
wave from the bounce at nuclear densities reaches its edge. The differences
in the electron fraction are of order \( 3 \)\%, the differences
in the neutrino abundances are even smaller.

\begin{figure*}
{\centering \resizebox*{1\textwidth}{!}{\includegraphics{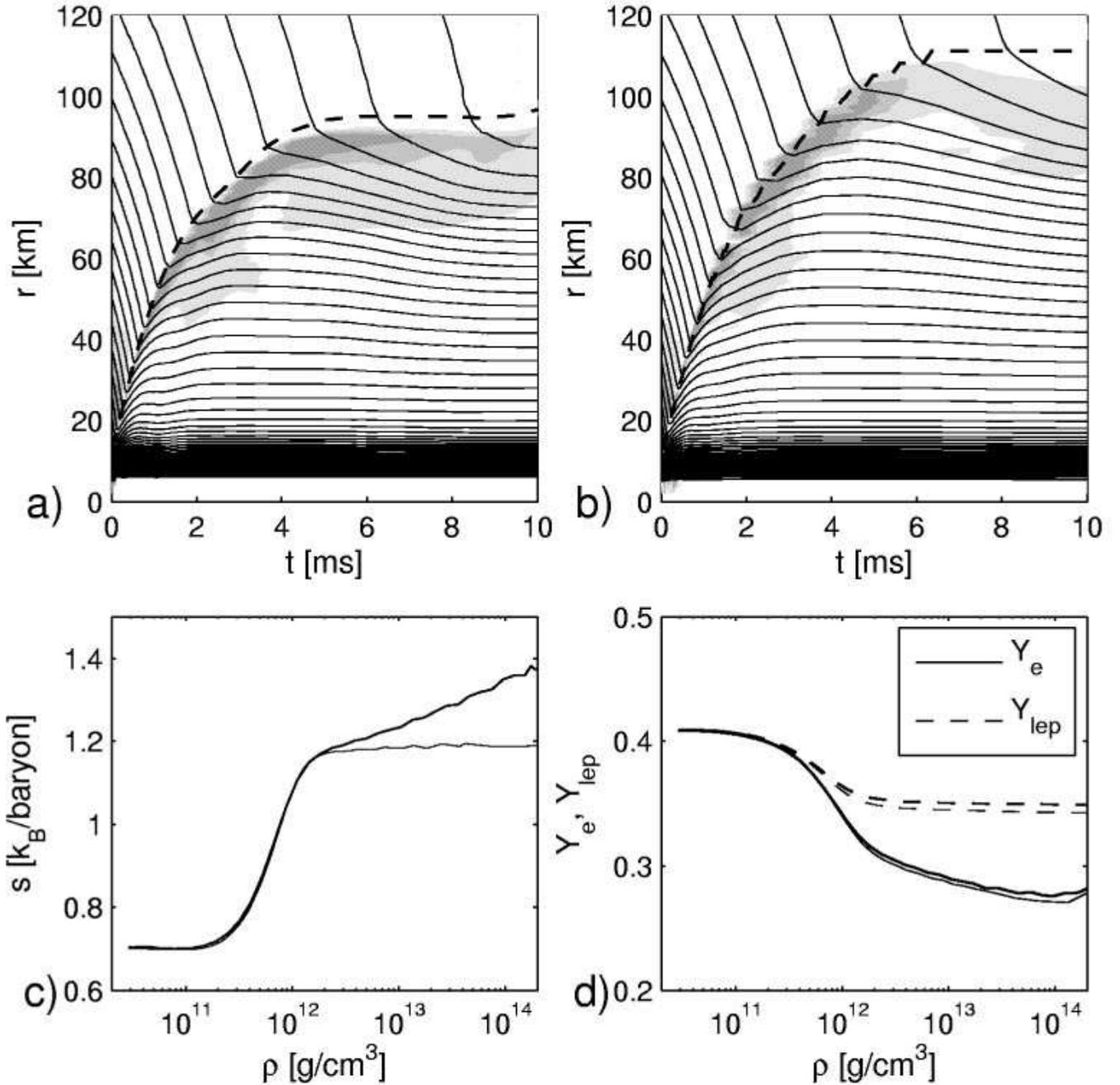}} \par}


\caption{Panel (a) shows the first \protect\( 10\protect \) ms after bounce
for model N13 in the simulation with \textsc{agile-boltztran}. Panel
(b) shows the same time period in the simulation with \textsc{vertex}.
The thin lines represent trajectories of fluid elements spaced with
an interval of \protect\( 0.02\protect \) M\protect\( _{\odot }\protect \).
The dashed line marks the shock position as a function of time. The
gradients of the mass trajectories at the right hand side of the dashed
line indicate the postshock velocities. At \protect\( \sim 4\protect \)
ms after bounce the shock has stalled. It turns into an expanding
accretion front with negative postshock velocities. Areas with strong
neutrino emission are shaded in three levels corresponding to values
of one, two, and three times \protect\( 10^{51}\protect \) neutrinos
per centimeter and second (i.e. for \protect\( 4\pi r^{2}\rho /m_{B}\times q_{\ell }\protect \),
where \protect\( q_{\ell }\protect \) is the lepton number source
term in units of leptons per baryon and second). The coincidence of
the launch of the neutrino burst with the transition from a dynamical
to an accretion shock in this model leads to an amplification of the
small hydrodynamics differences found above. The region behind the
weaker shock in \textsc{agile-boltztran} experiences more compression.
It therefore deleptonizes more rapidly and the weak shock loses even
more pressure support than the stronger shock in the \textsc{vertex}
simulation. The accretion front therefore expands more slowly in the
\textsc{agile-boltztran} simulation. Panel (c) compares the entropy
of the innermost zone as a function of density in \textsc{agile-boltztran}
(thick line) and in \textsc{vertex} (thin line). The agreement before
trapping is close to perfect, dynamically insignificant differences
appear at larger densities. Panel (d) shows an analogous comparison
for the electron fraction (solid lines) and lepton fraction (dashed
lines) in the innermost zone. \label{fig_A13}}
\end{figure*}

Most of these differences are introduced during the last \( 2 \)
ms before bounce. The profiles are in nearly perfect agreement before.
This becomes evident in panels (c) and (d) of Fig. \ref{fig_A13}
where we plot the entropy and lepton fractions of the innermost zone
as functions of density during core collapse. We find that the differences
in deleptonization appear just before neutrino trapping sets in, i.e.
when the effective electron capture rates are highest. The entropy
evolution shows perfect agreement during infall, but after neutrino
trapping, an unphysical entropy increase in the innermost zones takes
place in the \textsc{agile-boltztran} simulation. Fig. \ref{fig_B13_0}d
clearly demonstrates that this entropy increase only occurs in the
innermost zone. Probably more significant is the small deviation of
order \( 5\% \) in the lepton and electron fraction which appears
at the same time. It is not confined to the innermost zone and influences
the enclosed mass at shock formation.

The formation of the shock is visible in the velocity profiles in
panel (b) of Fig. \ref{fig_B13_0}. The difference in enclosed mass
\( \sim 3\% \) between \textsc{agile-boltztran} and \textsc{vertex}
is qualitatively consistent with the differences in the electron fraction
profiles in Fig. \ref{fig_B13_0}d. The infall velocities in the outer
core agree well. Panel (b) also shows the density profiles of the
N13 run at bounce. We conclude the discussion of bounce with the observation
that there are small visible differences between the two N13 simulations,
but none of them is likely to induce large deviations in the postbounce
evolution.

\begin{figure*}
{\centering \resizebox*{1\textwidth}{!}{\includegraphics{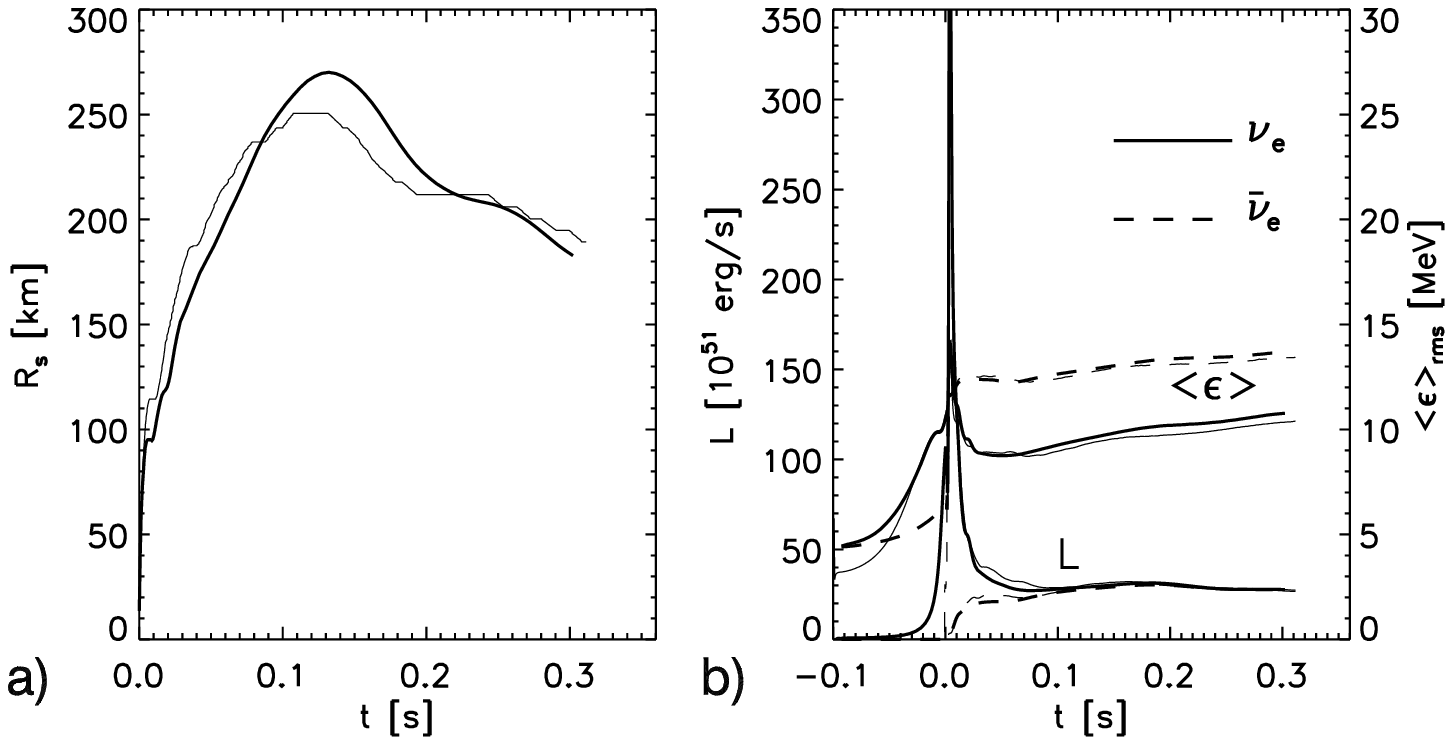}} \par}

\caption{The shock position as a function of time for model N13 is shown in
panel (a). The shock in \textsc{vertex} (thin line) propagates initially
faster and nicely converges after its maximum expansion to the position
of the shock in \textsc{agile-boltztran} (thick line). The neutrino
luminosities and rms energies for model N13 are presented as functions
of time in panel (b). The values are sampled at a radius of \protect\( 500\protect \)
km in the comoving frame. The solid lines belong to electron neutrinos
and the dashed lines to electron antineutrinos. The line width distinguishes
between the results from \textsc{agile-boltztran} and \textsc{vertex}
in the same way as in panel (a). The luminosity peaks are nearly identical,
the rms energies have the tendency to be larger in \textsc{agile-boltztran}.\label{fig_C13}}
\end{figure*}

Figure \ref{fig_A13}
shows the mass trajectories for both runs during the first \( 10 \)
ms after bounce, in panel (a) for \textsc{agile-boltztran} and in
panel (b) for \textsc{vertex}. The rising dashed line marks the shock
position. A first inspection reveals a difference of \( 15\% \) in
shock radius at \( 7 \) ms after bounce. As we will see later in
Fig. \ref{fig_C13}, this difference is transient. It appears after
a short dynamical phase of shock propagation, long before any neutrino
heating takes place and long before the accretion front has reached
its maximum radius at \( \sim 250 \) km in this optimistic model
with reduced input physics.

The discrepancy originates from a different shock strength as cause
and a different neutrino burst behavior as consequence and amplification
mechanism. The gradients of the mass trajectories in Figs. \ref{fig_A13}a
and b indicate the velocity of the material in the postshock region
(at the right hand side of the thick dashed line that represents the
position of the shock). The bounce-shock is dynamical at the beginning
and drives matter outward with positive postshock velocities. At about
\( 4 \) ms after bounce, the shock stalls because of the nuclear
dissociation of infalling matter and neutrino losses. It converts
into an accretion shock, characterized by jump conditions that connect
the high speed/low density infalling material to the low speed/high
density postshock material. There is an important difference between
the dynamical and the accretion shock: The postshock material behind
a dynamical shock is diluted between the rarefaction wave and the
shock front, while the material behind an accretion front continues
to be compressed due to the accumulated mass. The examination of the
mass trajectories in Fig. \ref{fig_A13}ab indicates that \textsc{vertex}
maintains a dynamical shock for a longer time than \textsc{agile-boltztran}.
This is consistent with the result of the hydrodynamics comparison
in Fig. \ref{fig_H13}.

This difference in shock propagation is initially not very large.
It is significantly amplified by the coincidence of the electron neutrino
burst with the transition from a dynamical to an accretion shock.
As the shock compresses the infalling lepton-rich material, the fermionic
electrons have to populate high energy levels and are rapidly converted
to neutrinos by captures on protons as soon as the density is low
enough for the produced neutrinos to escape. This neutrino burst can
be launched by the shock while it is still in its dynamical phase
or after it has stalled to an accretion front. If the neutrino burst
is launched during the dynamical phase, infalling matter deleptonizes
due to the immediate compression in the shock front. After that, electron
captures are reduced quickly because the matter re-expands behind
the dynamical shock and the density drops again. In an accretion shock,
the infalling matter experiences the same initial deleptonization
in the shock front, but continues to emit neutrinos due to the compression
behind the shock. The neutrino burst from an accretion shock is therefore
more intense. The neutrino emission behind the shock removes energy
and lepton number and reduces the pressure support. It accelerates
the attenuation of the weaker shock and thereby amplifies the initial
difference in shock strength. Using the lepton number source term,
\( q_{\ell } \), in units of generated leptons per baryon and second,
we shaded in panels (a) and (b) of Fig. \ref{fig_A13} the regions
where the neutrino emission \( 4\pi r^{2}\rho /m_{B}\times q_{\ell } \)
exceeds the thresholds of \( 10^{51} \) cm\( ^{-1} \)s\( ^{-1} \),
\( 2\times 10^{51} \) cm\( ^{-1} \)s\( ^{-1} \), and \( 3\times 10^{51} \)
cm\( ^{-1} \)s\( ^{-1} \). The comparison of panels (a) and (b)
illustrates that more neutrinos are lost from the region behind the
weaker shock in the \textsc{agile-boltztran} simulation. The initially
stronger shock in the \textsc{vertex} simulation suffers less neutrino
losses and leads to a more rapidly expanding accretion front.

\begin{figure*}
{\centering \resizebox*{1\textwidth}{!}{\includegraphics{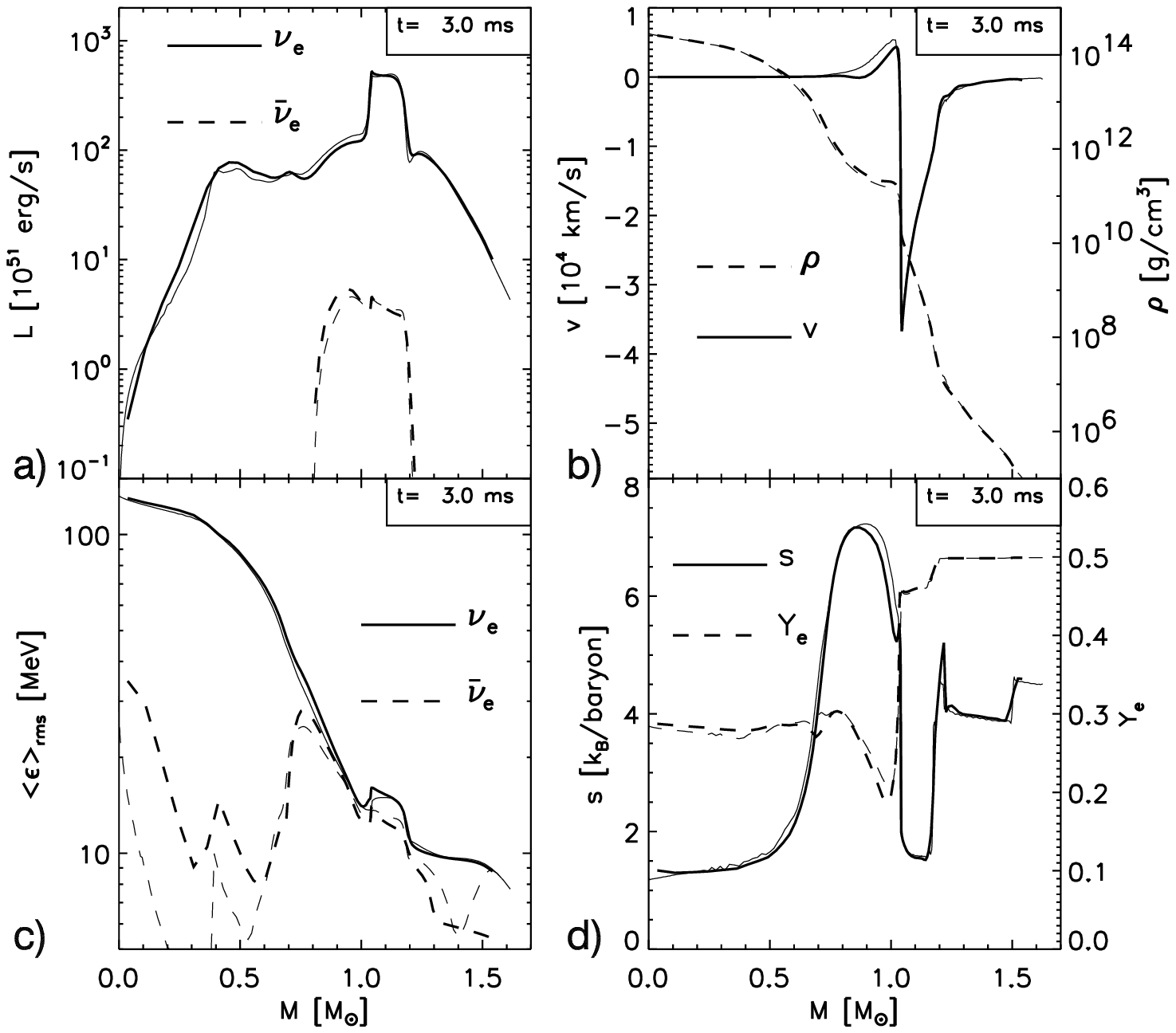}} \par}

\caption{Snapshots at \protect\( 3\protect \protect \) ms after bounce for
model N13. Data from the \textsc{agile-boltztran} simulation are drawn
with thick lines. Data from the \textsc{vertex} simulation are drawn
with thin lines. Panel (b) shows the velocity (solid lines) and density
(dashed lines) profiles, panel (d) the entropy (solid lines) and electron
fraction (dashed lines). Panels (a) and (c) show the neutrino luminosities
and rms energies respectively. Solid lines refer to electron neutrinos
and dashed lines to electron antineutrinos. The snapshot reveals similar
small differences as the hydrodynamics comparison in Fig. \ref{fig_H13}.\label{fig_B13_1}}
\end{figure*}

After this investigation we can easily interpret the time slice at
\( 3 \) ms after bounce presented in Fig. \ref{fig_B13_1}.
In the luminosity, panel (a), we see good agreement during the launch
of the neutrino burst. The somewhat higher rms energies of the burst
electron neutrinos, compared to the previously emitted ones, are visible
in panel (c). The entropy profile in panel (d) shows a slightly weaker
shock in \textsc{agile-boltztran} than in \textsc{vertex}, very similar
to the differences found for the hydrodynamics comparison in Fig.
\ref{fig_H13}. The electron fraction profiles now show that the deleptonization
of the postshock region between \( 0.8 \) and \( 1.0 \) M\( _{\odot } \)
is more pronounced in \textsc{agile-boltztran.} This is consistent
with the higher density visible in panel (b). The positive velocities
behind the shock also demonstrate that the \textsc{vertex} shock has
still a larger kinetic energy in this snapshot. The shock in \textsc{vertex}
will stall somewhat later than in \textsc{agile-boltztran}.

\begin{figure*}
{\centering \resizebox*{1\textwidth}{!}{\includegraphics{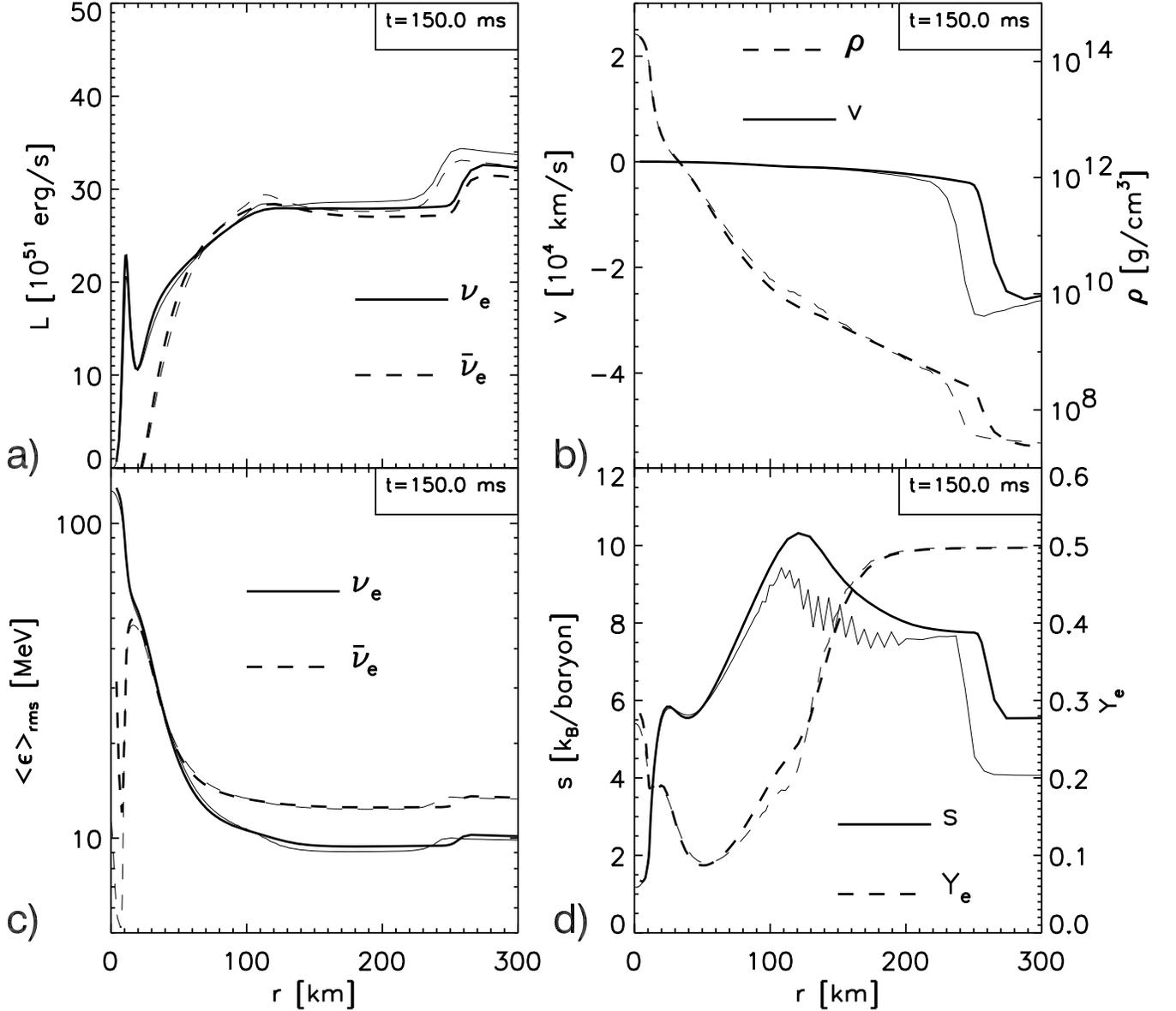}} \par}

\caption{Snapshots at \protect\( 150\protect \) ms after bounce for model
N13. Data from the \textsc{agile-boltztran} simulation are drawn with
thick lines. Data from the \textsc{vertex} simulation are drawn with
thin lines. Panel (b) shows the velocity (solid lines) and density
(dashed lines) profiles, panel (d) the entropy (solid lines) and electron
fraction (dashed lines). The neutrino luminosities and rms energies
are shown in panels (a) and (c) respectively. Solid lines refer to
electron neutrinos and dashed lines to electron antineutrinos. This
stationary-state situation is typical of the neutrino heating phase
at later time after bounce. The agreement is satisfactorily close
in most quantities.\label{fig_13B_2}}
\end{figure*}

The further evolution is best followed in the time-dependent diagrams
in Fig. \ref{fig_C13}.
In panel (b), the luminosities and rms energies are shown, sampled
at a fixed radius of \( 500 \) km. The two neutrino signals are qualitatively
very similar. The neutrino burst in {\sc agile-boltztran} is somewhat
broader than in {\sc vertex} and has a \( 7\% \) smaller peak luminosity.
The deviations after the burst are at most \( 15 \)\% around \( 50 \)
ms after bounce. This difference is a late consequence of the deleptonization
differences during the neutrino burst. The material in \textsc{vertex}
is left with higher electron fraction and higher entropy after the
neutrino burst. Hence it deleptonizes at a higher rate afterwards.
The rms energies tend to be lower in \textsc{vertex} than in \textsc{agile-boltztran}.
Finally, panel (a) compares the accretion front trajectories over
a longer period of time. We find the described differences in the
early expansion of the accretion front. At \( 85 \) ms after bounce,
however, the trajectories cross and a larger maximum radius is reached
in the {\sc agile-boltztran} simulation. The maximum radius of the
accretion front differs by \( \sim 8\% \) between the two simulations.
This difference is due to the higher preshock entropies in the {\sc agile-boltztran}
simulation visible in Fig. \ref{fig_13B_2}d. The difference in the
entropies of the infalling material stems from the interface between
the silicon layer and the material in nuclear statistical equilibrium
(NSE). The burning in {\sc agile-boltztran} cannot be switched off
completely for this comparison, because conversions between non-NSE
and NSE are unavoidable when the adaptive grid moves its zones relative
to the mass coordinate. This produces a local entropy peak at the
composition interface (see Fig. \ref{fig_B13_1}d at an enclosed mass
of \( 1.2 \) M\( _{\odot } \)) which crosses the accretion front
at \( 85 \) ms after bounce. The higher entropy leads to a temporarily
lowered accretion rate which allows the accretion front in the {\sc agile-boltztran}
simulation to propagate to a larger radius than in the {\sc vertex}
simulation where no conversions between non-NSE and NSE occur during
this simulation with reduced input physics. After the initial expansion
phase, where matter piles up on the neutron star hydrostatically,
the pressure support in the cooling region starts to diminish rapidly
and the matter in the heating region is pulled inward from below (see
e.g. \citet{Janka_01,Liebendoerfer_et_al_01}). Simultaneously, the
mass shell in the preshock region containing the described entropy
differences has fallen through the shock and the entropy becomes more
similar again. Therefore the trajectories of the accretion front converge
and agree well during the shock recession phase.

We finish the comparison of the N13 model with a closer look at a
time slice at \( 150 \) ms after bounce (Fig. \ref{fig_13B_2}),
which is a snapshot during this quasi-stationary accretion phase.
Panel (a) demonstrates excellent agreement of the luminosities in
all regions of the computational domain. The rms energies in panel
(c) show small differences, especially outside of \( 100 \) km radius.
The velocity profiles in panel (b) agree well if one disregards the
different shock positions explained above. The higher accretion rates
in {\sc vertex} of material with lower entropy lead to a higher density
in the cooling region. This is visible in the density profiles in
panel (b) and the entropy profiles in panel (d). The reaction time
scale is comparable to the dynamical time scale inside the gain radius
at \( 115-120 \) km. The infalling fluid element therefore is close
to weak equilibrium in the given neutrino background. Since the neutrino
luminosities are very similar in the two simulations, the larger density
of the \textsc{vertex} run requires a lower equilibrium electron fraction
in the cooling region inside the gain radius. The corresponding differences
in the pressure profiles imply less support for the shock and are
consistent with a smaller radius of the accretion front.

\subsection{General Relativistic 15 M\protect\( _{\odot }\protect \) Model}

\label{section_G15}
\begin{figure*}
{\centering \resizebox*{1\textwidth}{!}{\includegraphics{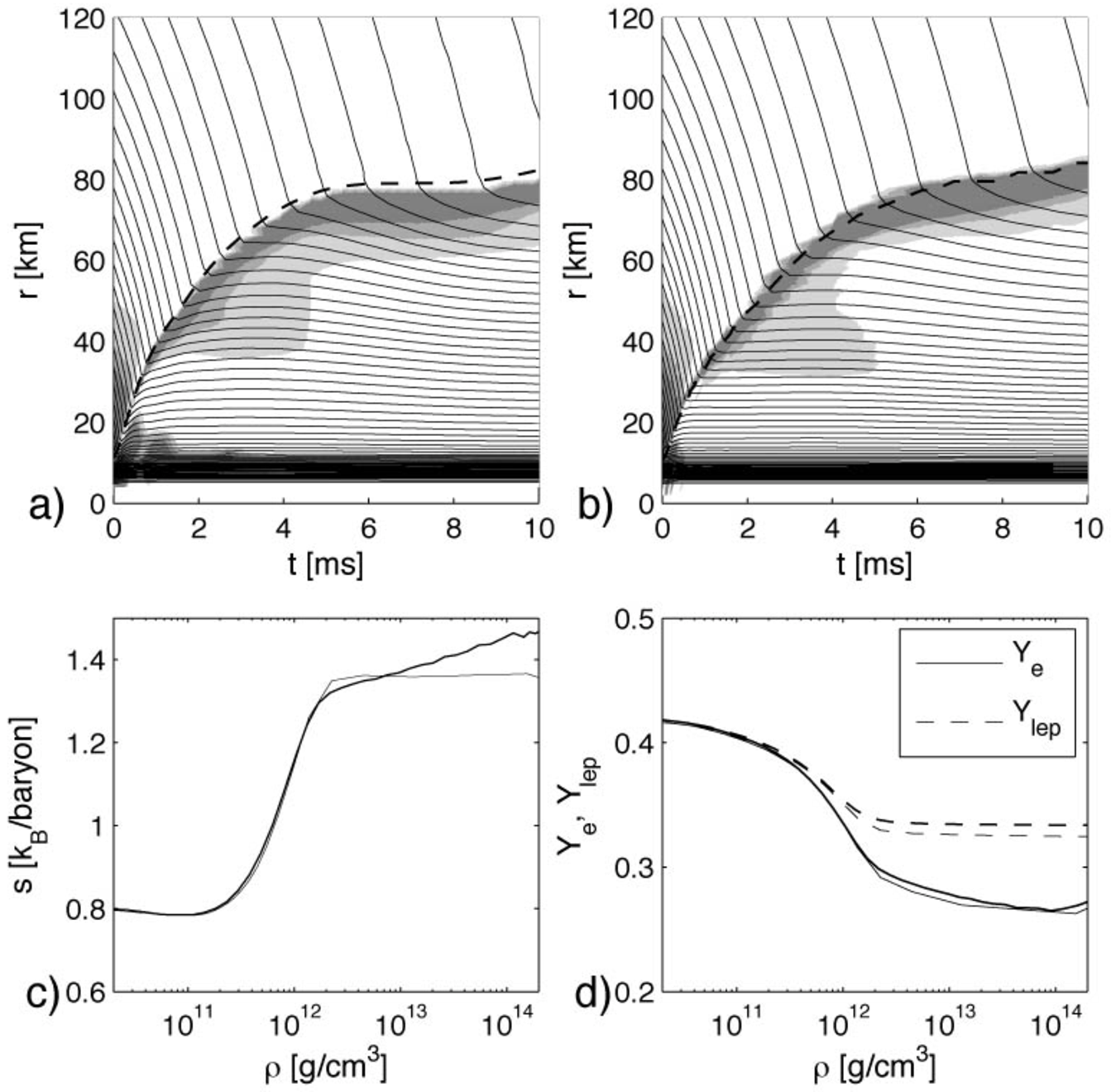}} \par}

\caption{Panel (a) shows the first \protect\( 10\protect \) ms after bounce
for model G15 in the simulation with \textsc{agile-boltztran}. Panel
(b) shows the same time period in the simulation with \textsc{vertex}.
The thin lines represent trajectories of fluid elements spaced with
an interval of \protect\( 0.02\protect \) M\protect\( _{\odot }\protect \).
Areas with strong neutrino emission are shaded in three levels corresponding
to values of one, two, and three times \protect\( 10^{51}\protect \)
neutrinos per centimeter and second (i.e. for \protect\( 4\pi r^{2}\rho /m_{B}\times q_{\ell }\protect \),
where \protect\( q_{\ell }\protect \) is the lepton number source
term in units of leptons per baryon and second). Both codes obtain
an extended region of strong neutrino emission behind the shock, which
turns into an accretion front (dashed line) at \protect\( 3-4\protect \)
ms after bounce. Panel (c) compares the entropy of the innermost zone
as a function of density in \textsc{agile-boltztran} (thick line)
and in \textsc{vertex} (thin line). The agreement before trapping
is close to perfect, dynamically insignificant differences appear
at larger densities. Panel (d) shows a comparison of the electron
fraction (solid lines) and lepton fraction (dashed lines) in the innermost
zone. The deviations between the two codes are of order \protect\( 3\%\protect \).\label{fig_A15}}
\end{figure*}

\begin{figure*}
{\centering \resizebox*{1\textwidth}{!}{\includegraphics{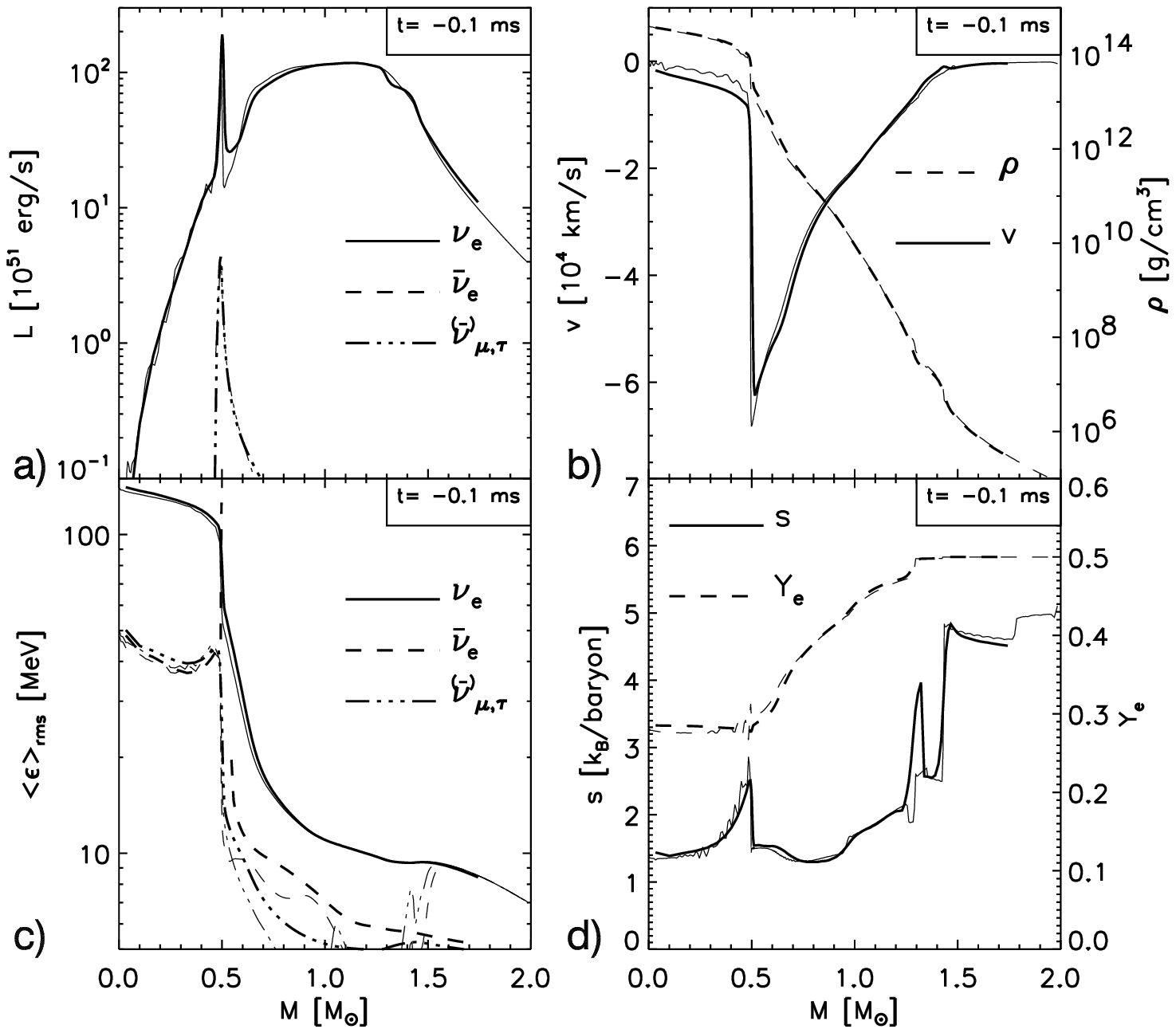}} \par}

\caption{Snapshots at bounce for model G15. Data from the \textsc{agile-boltztran}
simulation are drawn with thick lines. Data from the \textsc{vertex}
simulation are drawn with thin lines. Panel (b) shows the velocity
(solid lines) and density (dashed lines) profiles, panel (d) the entropy
(solid lines) and electron fraction (dashed lines). The neutrino luminosities
and rms energies are given in panels (a) and (c), respectively. Solid
lines correspond to electron neutrinos, dashed lines to electron antineutrinos,
and dash-dotted lines to \protect\( \mu \protect \)- or \protect\( \tau \protect \)-neutrinos
(or their antiparticles). In the region enclosed by the shock the
luminosities in \textsc{agile-boltztran} are smaller than in \textsc{vertex},
consistent with the larger central electron fraction and a slightly
larger enclosed mass at shock formation.\label{fig_B15_0}}
\end{figure*}

For the analysis of the G15 simulations we start
again with the description of the conditions at bounce. The evolution
of the entropy and lepton fraction in the innermost zone is shown
in panels (c) and (d) of Fig. \ref{fig_A15}.
Compared to model N13 with Newtonian gravity, the central entropy
in the general relativistic G15 model is \( 15\% \) higher and the
lepton and electron fractions are \( 5\% \) lower. The deviations
between the \textsc{vertex} and \textsc{agile-boltztran} simulations
are of order \( 3\% \) with the exception of an entropy increase
in the innermost zone in the \textsc{agile-boltztran} simulation.
As shown in Fig. \ref{fig_B15_0}d, excellent agreement is found in
all other regions of the model up to the burning front, where different
approximations in tracking the composition and nuclear burning explain
the larger differences. In contrast to the entropy difference in the
innermost zone, the small differences in the electron fraction apply
to the whole high-density region enclosed by the shock.

The luminosity profiles at bounce are displayed in Fig. \ref{fig_B15_0}a.
Outside of the shock front, they have been set during collapse and
agree well. Also the luminosities in the diffusive regime reveal no
mentionable differences. As a consequence of the differences of the
electron fraction visible in panel (d), the velocity and density profiles
in panel (b) show the formation of the shock front in \textsc{vertex}
at a slightly deeper point than in \textsc{agile-boltztran}. In consideration
of this displacement in the shock position, also the rms neutrino
energies in panel (c) are in satisfactory agreement.

The early postbounce evolution of the G15 model is less sensitive
to small differences than the previously discussed N13 simulation.
The main reason is the weaker bounce shock. General relativistic effects
during core collapse shift the sonic point to a \( 20\% \) smaller
enclosed mass and lead to a less energetic bounce shock \citep{Liebendoerfer_et_al_01, Bruenn_DeNisco_Mezzacappa_01}
which has to dissociate more infalling material. The inclusion of
\( \mu  \)- and \( \tau  \)-neutrinos in the G15 runs causes additional
energy drain from the region behind the shock. After the shock has
stalled within \( 1 \) ms in both G15 simulations, the electron neutrino
burst is launched during the accretion shock phase. As expected from
the discussion of the neutrino emission from the shock in the N13
model, panels (a) and (b) in Fig. \ref{fig_A15} reveal a well extended
region of high neutrino emission behind the shock. Since the postshock
matter develops negative velocities a few milliseconds after bounce,
the further evolution is determined by the continued accumulation
of accreted matter, more and more effectively cooled as the accretion
front reaches layers with lower matter densities and neutrino opacities.
This quasi-stationary evolution is less sensitive to differences in
the numerics or input physics than the dynamical shock propagation
in the more optimistic N13 simulation. The feedback between neutrino
transport and hydrodynamics amplifies differences in the propagation
of dynamical shocks, because the material behind the weaker shock
emits more neutrinos so that the shock loses even more pressure support.
In case of a quasi-stationary accretion front, differences are likely
to be reduced because a larger accretion rate produces larger neutrino
losses.

\begin{figure*}
{\centering \resizebox*{1\textwidth}{!}{\includegraphics{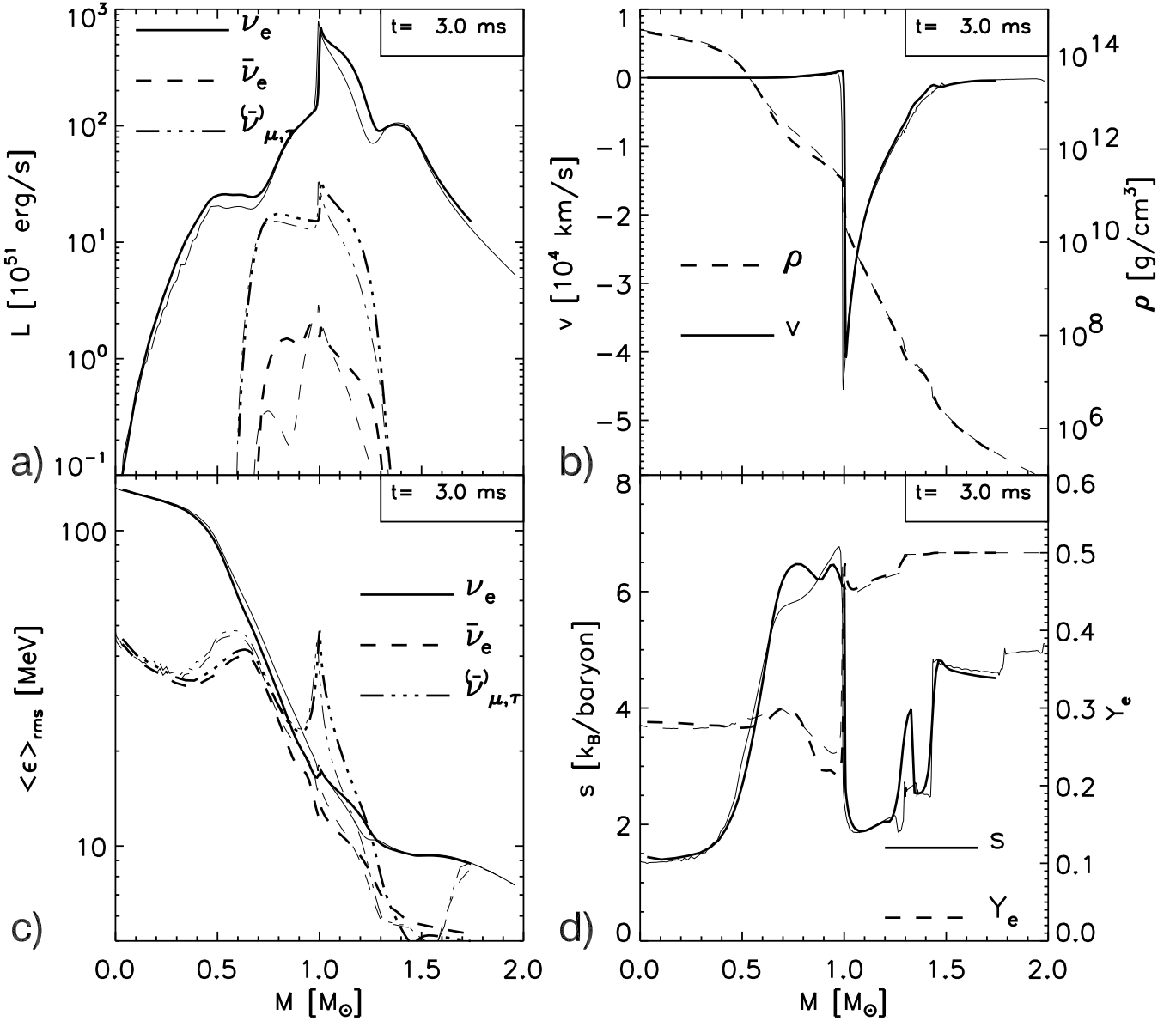}} \par}

\caption{Snapshots at \protect\( 3\protect \protect \) ms after bounce for
model G15. Data from the \textsc{agile-boltztran} simulation are drawn
with thick lines. Data from the \textsc{vertex} simulation are drawn
with thin lines. Panel (b) shows the velocity (solid lines) and density
(dashed lines) profiles, panel (d) the entropy (solid lines) and electron
fraction (dashed lines). The neutrino luminosities and rms energies
are given in panels (a) and (c), respectively. Solid lines correspond
to electron neutrinos, dashed lines to electron antineutrinos, and
dash-dotted lines to \protect\( \mu \protect \)- or \protect\( \tau \protect \)-neutrinos
(or their antiparticles). Variations in the entropy profiles reflect
the differences in the shock strength at bounce. The deleptonization
by the neutrino burst occurs an instant earlier in the \textsc{agile-boltztran}
simulation.\label{fig_B15_1}}
\end{figure*}

However, differences can still be observed: The \textsc{agile-boltztran}
shock is stronger at formation. As in the N13 simulations, this is
partly due to the higher electron fraction in the homologous core
of the \textsc{agile-boltztran} simulation. A difference in the entropy
profiles is left behind when the shock passes an enclosed mass of
\( 0.75 \) M\( _{\odot } \) in Fig. \ref{fig_B15_1}d. This points
to a larger deviation between the initial shock strengths than in
the N13 simulations so that the \textsc{agile-boltztran} shock expands
initially faster in Fig.~\ref{fig_A15}a. The differences between
the electron fraction profiles in Fig. \ref{fig_B15_1}d, however,
suggest that later on a faster deleptonization in \textsc{agile-boltztran}
damps the expansion of the accretion front and as a consequence the
shock positions in both simulations converge again. This connection
between a somewhat enhanced neutrino loss and a deceleration of the
expansion of the accretion front is supported by the shaded areas
in panels (a) and (b) of Fig. \ref{fig_A15}, which highlight differences
in the regions of strong neutrino emission in both simulations.

In Fig. \ref{fig_B15_1}b, a lower density at \( 0.75 \) M\( _{\odot } \)
is caused by the higher entropy in the \textsc{agile-boltztran} run.
The velocity profiles at \( 3 \) ms after bounce are in very good
agreement. A slightly higher infall velocity in front of the shock
leads to a positive entropy gradient between \( 0.8 \) M\( _{\odot } \)
and \( \sim 1 \) M\( _{\odot } \) in \textsc{vertex}. Also the neutrino
luminosities at \( 3 \) ms after bounce in Fig. \ref{fig_B15_1}a
do not reveal new features, except for the presence of \( \mu  \)-
and \( \tau  \)-neutrinos that had not been included in the N13 run.
Because of the absence of charged-current reactions of these neutrinos,
they decouple at higher densities and reach appreciable luminosities
earlier in the evolution, before the appearance of electron antineutrinos,
which are initially suppressed by the high electron degeneracy. The
apparent difference in the electron antineutrino luminosity between
\textsc{agile-boltztran} and \textsc{vertex} is due to a small time
lag in displaying this rapidly rising quantity.

\begin{figure*}
{\centering \resizebox*{1\textwidth}{!}{\includegraphics{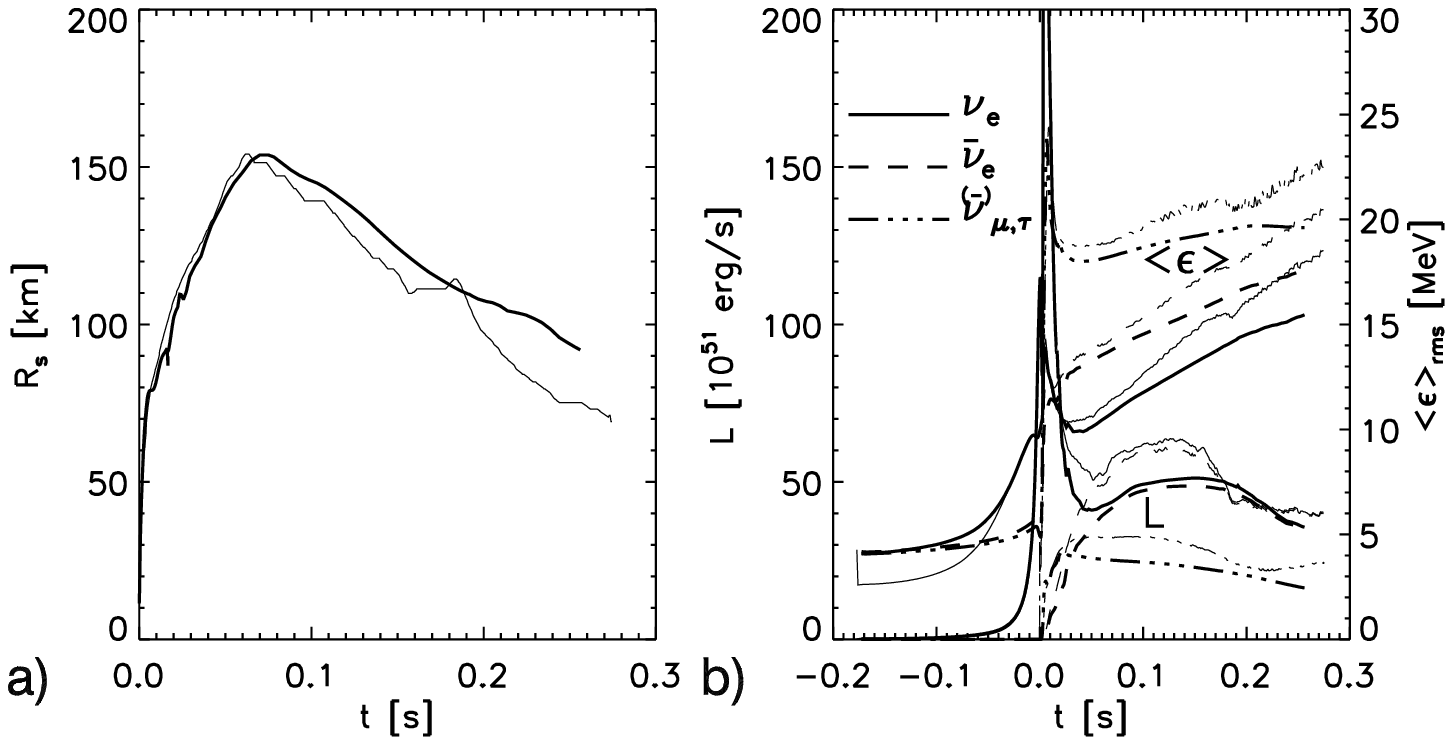}} \par}

\caption{Panel (a) shows the position of the accretion front as a function
of time for model G15. The two simulations predict a very similar
early expansion of the accretion front up to a maximum radius of \protect\( 150\protect \)
km. Then, the accretion front in \textsc{vertex} (thin line) retreats
faster than in \textsc{agile-boltztran} (thick line) in response to
the slightly stronger contraction of the nascent neutron star due
to the approximate treatment of general relativity in the \textsc{vertex}
simulation. The hump in the \textsc{vertex} shock position at \protect\( \sim 180\protect \)
ms after bounce corresponds to an entropy and density discontinuity
at the bottom of the oxygen-rich silicon shell. Because the \textsc{vertex}
simulation resolves the structure of the outer layers of the progenitor
star more accurately than the diffusive adaptive grid in \textsc{agile-boltztran},
this feature is less pronounced in the latter simulation. The neutrino
luminosities and rms energies are presented as functions of time in
panel (b). The values are sampled at a radius of \protect\( 500\protect \)
km in the comoving frame. The solid lines correspond to electron neutrinos,
the dashed lines to electron antineutrinos, and the dash-dotted lines
to \protect\( \mu \protect \)- or \protect\( \tau \protect \)-neutrinos
(or their antiparticles). The line width distinguishes between the
results from \textsc{agile-boltztran} (thick lines) and \textsc{vertex}
(thin lines). The differences in the neutrino results are mainly---but
not exclusively---indirect consequences (due to the more compact neutron
star) of the approximate treatment of general relativity in the \textsc{vertex}
simulation.\label{fig_C15}}
\end{figure*}

The further evolution is resumed in Fig. \ref{fig_C15}.
Panel (a) compares the position of the accretion front as a function
of time for the G15 models. During the expansion of the accretion
front, we find very good agreement and the maximum radius is nearly
identical. Afterwards, the accretion front retreats somewhat more
slowly in the \textsc{agile-boltztran} simulation than in the \textsc{vertex}
simulation. In the latter, the retraction transiently stagnates between
\( 150 \) and \( 180 \) ms after bounce. Such features are caused
by the steep density drop at the infalling interfaces between layers
of different composition outside of the iron core. The transition
to the oxygen-rich silicon layer passes the shock at about \( 165 \)
ms after bounce. \textsc{agile-boltztran} tracks the structure of
the outer layers less accurately because of the artificial diffusion
introduced by the adaptive grid. Discrete transitions between layers
are therefore washed out to some extent such that their impact on
the trajectory of the accretion front is less pronounced, although
still qualitatively visible. The luminosity peaks during the electron
neutrino burst deviate only by \( 3\% \) in the G15 simulations,
the average peak value is \( 3.8\times 10^{53} \) erg/s with a half-width
of \( 6 \) ms. The further time evolution of the luminosities and
rms energies in panel (b) reveals \( \sim 20\% \) larger values in
the \textsc{vertex} run. These are at least in part a consequence
of the increased accretion rate in the \textsc{vertex} simulation
during the retraction phase of the accretion front. We will make this
argument more precise in the following discussion of late time slices.

\begin{figure*}
{\centering \resizebox*{1\textwidth}{!}{\includegraphics{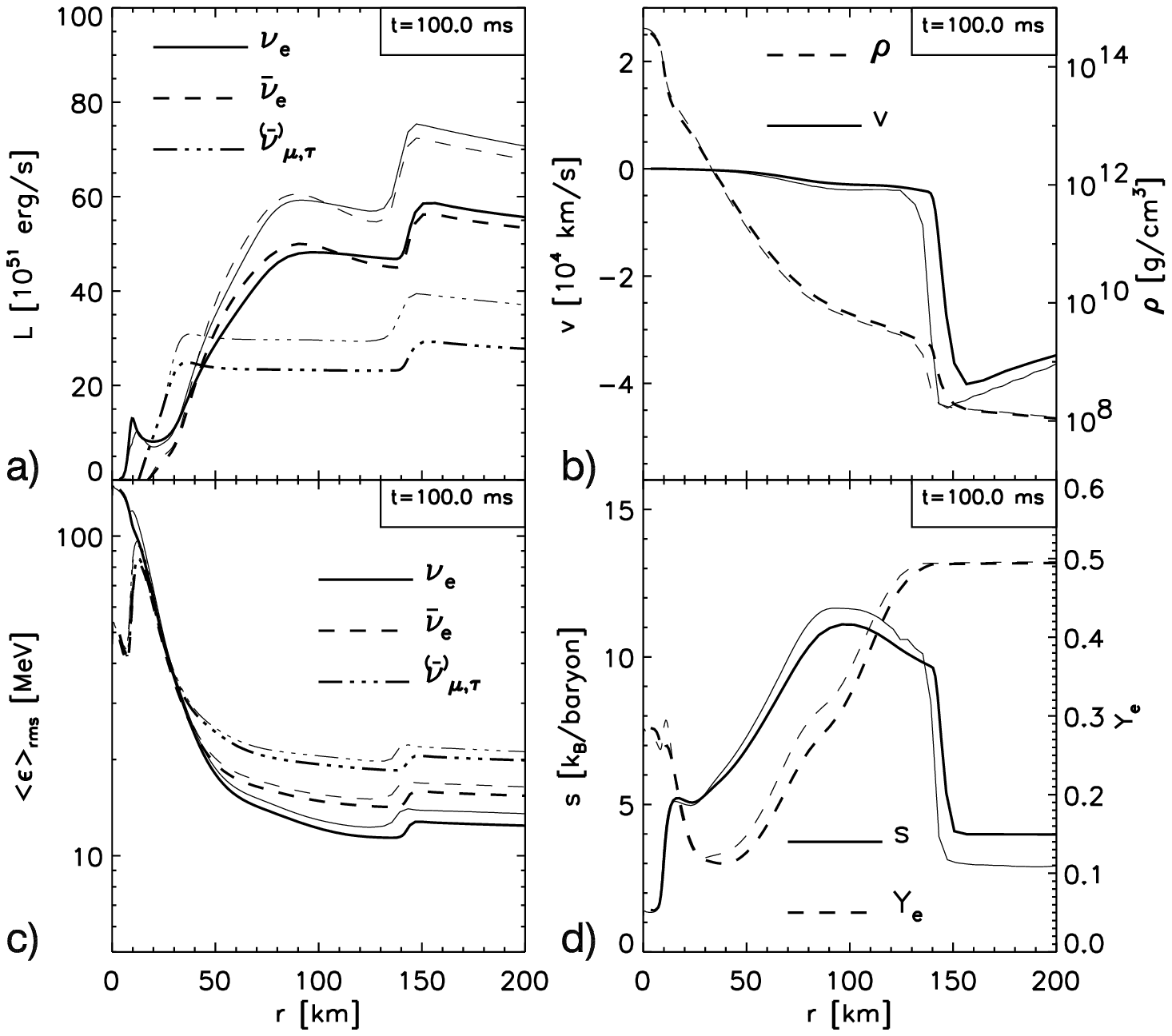}} \par}

\caption{Snapshots at \protect\( 100\protect \) ms after bounce for model
G15. Data from the \textsc{agile-boltztran} simulation are drawn with
thick lines. Data from the \textsc{vertex} simulation are drawn with
thin lines. Panel (b) shows the velocity (solid lines) and density
(dashed lines) profiles, panel (d) the entropy (solid lines) and electron
fraction (dashed line). The neutrino luminosities and rms energies
are displayed in panels (a) and (c), respectively. Solid lines correspond
to electron neutrinos, dashed lines to electron antineutrinos, and
dash-dotted lines to \protect\( \mu \protect \)- or \protect\( \tau \protect \)-neutrinos
(or their antiparticles). In this time slice we find a smaller shock
position, somewhat faster infall ahead of the shock, and higher postshock
infall velocities in \textsc{vertex} compared to the results of \textsc{agile-boltztran}.
Consistent with the more compact structure of the protoneutron star,
the entropies, neutrino luminosities, and neutrino rms energies in
\textsc{vertex} are larger than in \textsc{agile-boltztran}.\label{fig_B15_2}}
\end{figure*}

We present two time slices for the long-term evolution of the G15
simulation. The first time slice is at \( 100 \) ms after bounce
when the neutrino heating is most efficient. The second time slice
at \( 250 \) ms marks the end of the time period covered by the simulations.
The time slice at \( 100 \) ms is given in Fig. \ref{fig_B15_2}.
Panel (b) shows the shock in \textsc{vertex} at a slightly smaller
radius and the preshock infall velocities to be somewhat higher than
in \textsc{agile-boltztran}. The luminosities in panel (a) are up
to \( 20\% \) larger in \textsc{vertex}. The luminosity discontinuity
across the shock front caused by the Doppler frequency shift and angular
aberration for an observer in the comoving frame is also larger in
\textsc{vertex} because of the larger luminosity and the larger velocity
jump visible in panel (b). Otherwise the relative differences between
the two runs are just inverse to the situation we analyzed in Fig.
\ref{fig_13B_2} in case of the N13 simulation. Now \textsc{agile-boltztran}
has a higher density in the shocked material in panel (b) and a correspondingly
lower entropy and electron fraction in panel (d). Also the neutrino
rms energies in panel (c) are now lower than in the \textsc{vertex}
simulation.

\begin{figure*}
{\centering \resizebox*{1\textwidth}{!}{\includegraphics{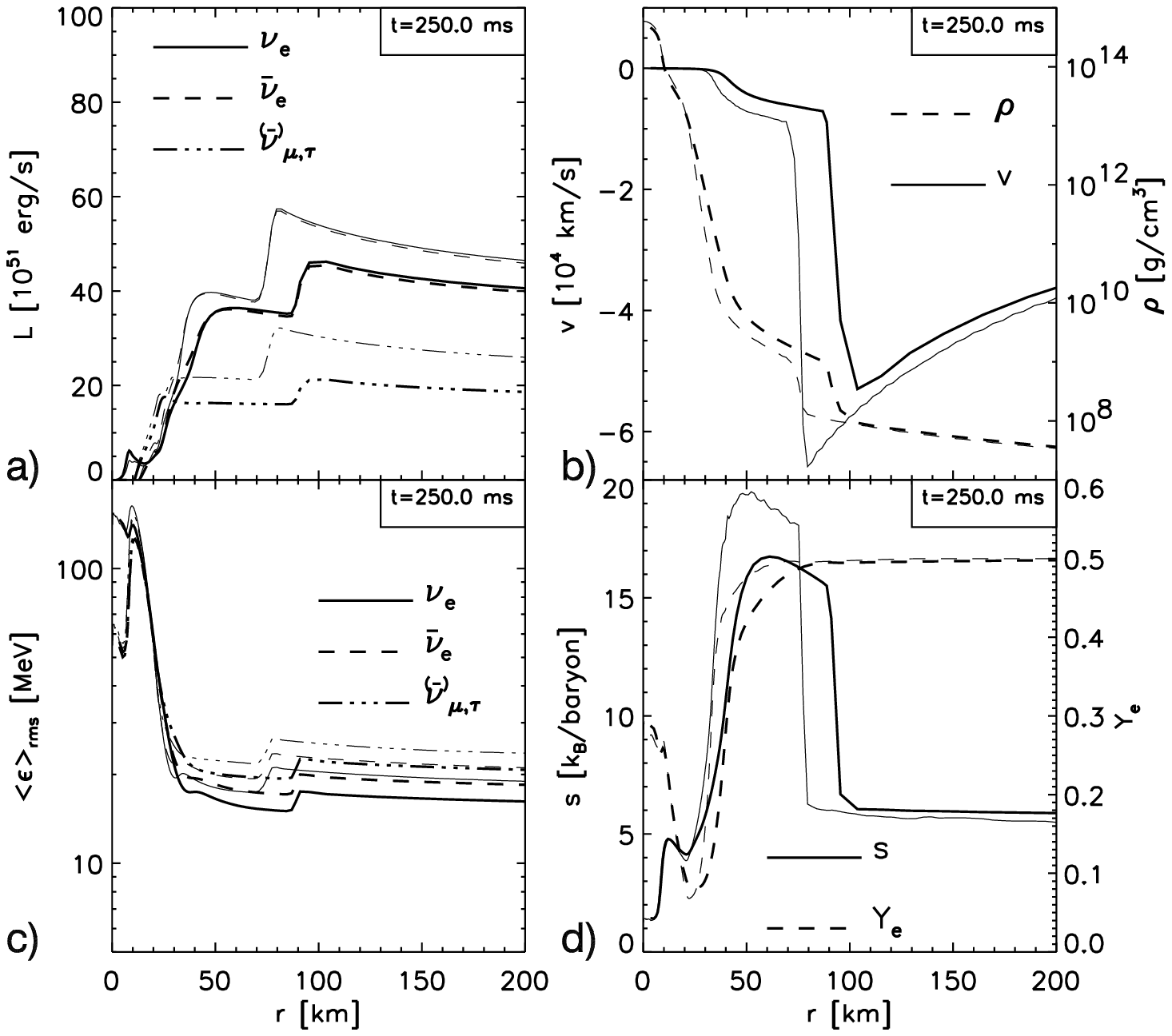}} \par}

\caption{Snapshots at \protect\( 250\protect \) ms after bounce for model
G15. Data from the \textsc{agile-boltztran} simulation are drawn with
thick lines. Data from the \textsc{vertex} simulation are drawn with
thin lines. Panel (b) shows the velocity (solid lines) and density
(dashed lines) profiles, panel (d) the entropy and electron abundances.
The neutrino luminosities and rms energies are displayed in panels
(a) and (c). Solid lines correspond to electron neutrinos, dashed
lines to electron antineutrinos, and dash-dotted lines to \protect\( \mu \protect \)-
or \protect\( \tau \protect \)-neutrinos (or their antiparticles).
The protoneutron star in the \textsc{vertex} simulation is more compact
than in the \textsc{agile-boltztran} simulation.\label{fig_B15_3}}
\end{figure*}

We also include the latest time slice at \( 250 \) ms after bounce
in Fig. \ref{fig_B15_3}. The panels show the same qualitative features
we have discussed in the context of Fig. \ref{fig_B15_2}, but at
this late time to a much larger extent. The protoneutron star in the
\textsc{vertex} simulation is more compact, causing the accretion
front to reside at a smaller radius and the luminosities of all neutrino
flavors to be larger and to have harder spectra than in the \textsc{agile-boltztran}
simulation. Due to the smaller radius of the accretion front and higher
infall velocities ahead of it, the postshock entropy is significantly
higher in the \textsc{vertex} run.

A more compact neutron star with a larger mass and a stronger gravitational
potential (which would lead to higher preshock velocities) can be
a consequence of a higher mass accretion rate. Indeed we observe differences
in the mass flux to the shock, which can be caused by the different
quality to follow structures in the outer stellar layers in both simulations.
An infalling density feature in \textsc{vertex} (which might evolve
differently due to the different treatment of {}``burning'' and
thus different entropy and pressure, or may be smoothed by the diffusivity
of the adaptive grid in \textsc{agile-boltztran}) transiently enhances
or reduces the accretion rate. When, for example, the transition to
the oxygen-rich silicon layer falls in shortly before \( 200 \) ms
after bounce, the accretion rate decreases sharply and the retraction
of the accretion front in the \textsc{vertex} simulation stagnates.
This is consistent with the luminosity reduction during this phase.
But the artificial diffusion in the outer layers of the \textsc{agile-boltztran}
simulation can explain only transient differences of the mass accretion
rate and of the total accreted mass, because an associated redistribution
of matter and a modification of the preshock structure is limited
to a certain radial domain. It should, however, not produce persistent
differences in the density distribution behind the shock, which in
fact can be seen in Figs.~\ref{fig_B15_2} and \ref{fig_B15_3} (cf.\ the
density profiles in panels (b)). A closer inspection of our models
in fact reveals that the mass accretion rate outside of the shock
and the baryonic mass accumulated in the neutron star show temporary
differences only between \( \sim  \)30\( \,  \)ms and \( \sim  \)200\( \,  \)ms,
but become very similar again toward the end of our simulations.

\begin{figure}
{\centering \resizebox*{0.5\textwidth}{!}{\includegraphics{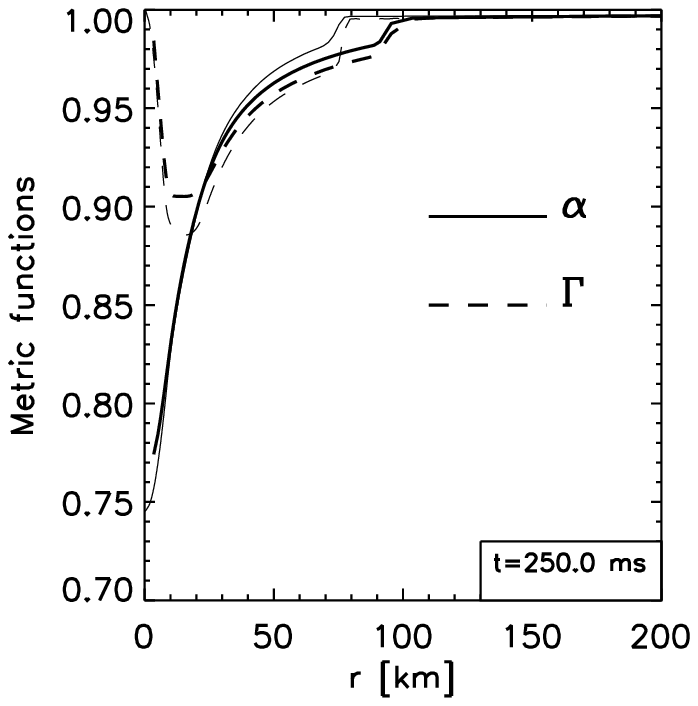}} \par}

\caption{Metric coefficients \protect\( \alpha =g_{tt}^{1/2}\protect \) and
\protect\( \Gamma =g_{aa}^{-1/2}\partial r/\partial a=\left[ 1+(u/c)^{2}-2Gm/(rc^{2})\right] ^{1/2}\protect \)
in the G15 run at \protect\( 250\protect \) ms after bounce. The
discontinuity of the lapse function across the shock front reflects
the fact that the comoving frame is not an inertial system in the
special relativistic limit. Results from {\sc agile-boltztran} are
plotted with tick lines, those from {\sc vertex} with thin lines.
The higher compactness of the neutron star in the {\sc vertex} run
is obvious. Ahead of the shock metric effects are very small.\label{fig_D15}}
\end{figure}

The systematically evolving and growing difference during the long-term
evolution must therefore be caused by another effect. The combination
of higher luminosities, higher rms energies and higher entropies at
the neutrinosphere reminds us of the differences found between Newtonian
and general relativistic simulations, where they are due to differences
between the Newtonian and relativistic gravitational potential \citep{Bruenn_DeNisco_Mezzacappa_01,Liebendoerfer_et_al_01}.
Both of our numerical methods are designed to accurately describe
the hydrostatic structure of the protoneutron star according to the
solution of the Tolman-Oppenheimer-Volkoff (TOV) equation. But in
the relativistic case differences in the potential can not only be
caused by a difference of the enclosed mass at a given radius. In
contrast to the Newtonian case, the relativistic potential depends
highly nonlinearly on the structure of the configuration through its
dependence on the mass distribution, pressure, and energy density.
The gravitational potential is therefore sensitive to differences
in the early post-bounce dynamics of the propagating shock (e.g.,
due to the different initial shock strength) and to pressure and entropy
differences created at later times, e.g., associated with the transient
differences of the mass accretion rate or due to the higher infall
velocities ahead of the shock in the \textsc{vertex} run. The overestimation
of the velocities in the collapse layer is also a consequence of the
approximation of general relativity in case of \textsc{vertex}. The
latter code uses only a gravitational potential that is corrected
for general relativistic effects, but ignores relativistic kinematics.
When the preshock values of the infall velocities reach 10--15\% of
the speed of light, the velocities computed by \textsc{vertex} are
overestimated in comparison to the relativistic velocities calculated
by \textsc{agile-boltztran}. This again has an influence on the long-term
post-bounce evolution and thus feeds back into the core structure
and causes a nonlinear response of the relativistic potential. While
we find that the maximum densities at bounce and the following relaxation
to a static situation are in good agreement between the \textsc{vertex}
and \textsc{agile-boltztran} runs, we subsequently observe clear deviations
of the central density and of the density profile which gradually
evolve and grow at later times. The \textsc{vertex} simulation develops
a higher central density and a steeper density gradient outside of
the high-density core, and thus a more compact neutron star with a
higher relativistic potential. This is confirmed by Fig. \ref{fig_D15}
which shows the profiles of the metric coefficients \( \alpha =g_{tt}^{1/2} \)
and \( \Gamma =g_{aa}^{-1/2}\partial r/\partial a \) in Eq. (\ref{eq_comoving_metric})
at \( 250 \) ms after bounce. The smaller deviations from unity of
the metric coefficients in the {\sc agile-boltztran} run are consistent
with the less compact structure of the protoneutron star in this simulation.
The profiles also show that the metric coefficients are nearly unity
outside of the accretion front. The larger preshock velocities for
the {\sc vertex} run in Figs. \ref{fig_B15_2}b and \ref{fig_B15_3}b
are therefore not a consequence of the fact that the lapse function
is not included in the velocity plotted for this simulation. They
are more likely caused by the disregard of kinematical effects and
a stronger gravitational potential in the approximation of general
relativity used in {\sc vertex}.

\subsection{Discussion}

\label{section_discussion}This work extends the testing of our codes
beyond the independently performed calculations of idealized problems
which have analytical solutions \citep{Messer_00, Rampp_Janka_02, Liebendoerfer_Rosswog_Thielemann_02, Liebendoerfer_et_al_04}.
Here we directly compare the results of the codes in the application
they were actually developed for. Our aim was to assess quantitative
differences in complex supernova simulations to reduce the probability
of qualitative differences in future applications. We also intended
to create points of reference for future testing of codes that handle
the challenges of supernova physics, and to lay the foundations for
performing such tests in a more realistic way than by means of comparison
with analytic solutions of idealized test problems.

We have encountered two fundamental difficulties in our comparison.
The first is the fact that the two codes employ different methods,
use different basic quantities, and are differently structured. The
comparison of the results of the two approaches is straightforward,
but it is by far more challenging (and sometimes impossible) to track
differences back to their origin if the compared quantities are not
calculated in a similar way. The second difficulty is more related
to supernova physics than to the methodology. In the comparison of
our results we have very often encountered the situation that all
quantities within either one simulation are perfectly consistent,
but still not the same as in the other simulation. Further investigations
of the differences revealed small initial differences in several tightly
coupled quantities that grow with ongoing evolution. Because of the
strong feedbacks in the supernova problem it was often almost impossible
to separate cause and consequences of the deviations. This strong
coupling between quantities indicates that the problem is governed
by many equilibria. Sometimes the results converge again and find
back to a similar evolution as soon as the equilibrium is achieved.
This is for example the case in model N13 when stationary-state accretion
is established after the very dynamical early postbounce phase.

When we consider just the overall dynamical evolution of the spherically
symmetric postbounce phases in Figs. \ref{fig_C13} and \ref{fig_C15},
we find agreement in all qualitative features of the history. But
we also find some significant quantitative differences. For example,
the early shock propagation and the luminosities during the first
\( 100 \) ms after bounce are different in model N13. We attribute
an important part of these differences to the hydrodynamical shock
propagation that appears to maintain a stronger shock in \textsc{vertex}
than in \textsc{agile-boltztran}. The close coincidence of the neutrino
burst with the transition from a dynamical to an accretion shock in
this model leads to an amplification of existing differences. The
region behind the weaker shock in \textsc{agile-boltztran} deleptonizes
to a larger extent than behind the slightly stronger shock in \textsc{vertex}.
While the deleptonization burst is more extended in {\sc agile-boltztran},
the \textsc{vertex} core could maintain higher luminosities later
on. This may be the reason for a somewhat more optimistic shock propagation
in the early phase of the \textsc{vertex} simulation of model N13.

No such amplification effect takes place in the G15 model where the
shocks in both runs have already made the transition to an accretion
front when they reach densities from where neutrinos begin to escape.
Despite of some differences in the time-dependence of the shock strength,
the evolution of these two more realistic runs with {}``standard''
input physics agrees nicely during the early post-bounce phase. We
find good agreement of the neutrino quantities in the diffusive inner
core of the protoneutron star and the timing and peak height of the
neutrino burst. The approximations of general relativistic effects
in \textsc{vertex} yield accurate results until bounce and do not
introduce larger uncertainties with respect to the general relativistic
approach than other acceptable approximations. Differences appear
only in the later evolution when the outer layers of the progenitor
fall into the stalled accretion front. Some of these differences are
caused by a different description of nuclear burning in both codes
and a different capability to track the composition interfaces in
the outer layers of the collapsing core. The main reason for the differences,
however, is the approximate treatment of general relativity in the
\textsc{vertex} simulation. The basically Newtonian hydrodynamics
code employs a relativistically modified gravitational potential which
in principle allows one to accurately describe hydrostatic configurations
according to the TOV equation \citep{Rampp_Janka_02}. But the code
disregards the effects of relativistic kinematics which causes an
overestimation of the infall velocity ahead of the shock. The corresponding
differences in the dynamical evolution feed back into the gravitational
potential in a nonlinear way. This leads to a slightly more compact
neutron star, a somewhat smaller radius of the accretion front, and
a faster infall of matter between shock and neutron star, producing
up to \( \sim 20 \)\% higher accretion luminosities and rms energies
of neutrinos and antineutrinos of all flavors. While most of these
discrepancies in the neutrino quantities are a consequence of the
different structure of the accretion layer in the \textsc{vertex}
and \textsc{agile-boltztran} simulations, a smaller contribution may
also be ascribed to the fact that \textsc{vertex} takes into account
general relativistic redshift, but ignores the metric effects in the
radial coordinate.

The overall evolution, however, is consistent between both runs also
in case of the relativistic G15 model. Even more, both computations
produce not only a remarkable qualitative similarity of the behavior
during all phases but also show nice agreement in most features of
the radial profiles of the important quantities. This result may be
especially useful for multi-dimensional simulations, where essential
general relativistic effects should not be ignored, but a full relativistic
treatment might not have highest priority.

Both methods have their vulnerabilities and some of them have led
to lively discussions in the past. \textsc{agile-boltztran} was criticized
because of the rigorous approach and the generous consumption of computer
memory and CPU time. The resolution of the neutrino phase space was
considered to be at the lower justifiable limit in earlier simulations.
And indeed, the certainty that the evolution follows basic physical
principles independently of the resolution must be earned by very
specific twists and wrinkles in the finite difference representation
\citep{Liebendoerfer_et_al_04}. \textsc{vertex} has raised concerns
with regard to consistency as it uses disjunct gridding for radiation
transport and hydrodynamics and applies regridding procedures after
bounce. The separate solution of the transport equations for neutrino
number and neutrino energy only adds to the complexity. However, we
did not find discrepancies in our comparison that would support any
of these concerns to a degree that would question the reliability
of the qualitative results of our explosion-free supernova simulations.

The different strengths of the codes are more visible in the quantitative
details of the calculations. \textsc{vertex} produces great angular
resolution in the flux factors even far from the neutrinospheres and
keeps properly track of the sharp discontinuities in the composition
of the outer layers. Its extendibility to two-dimensional simulations
is built-in \citep{Rampp_Janka_02} and the general relativistic approximation
can be expected to produce good results in multi-dimensional situations
as well \citep{Buras_et_al_03}. However, the solution of a model Boltzmann
equation is much more involved in more than one dimension. Additional
approximations introduced by spherical averaging or ray-by-ray techniques
cannot be tested in a comparison between spherically symmetric models.
If these approximations are good, the variable Eddington factor approach
is a very efficient technique for multidimensional simulations.

\textsc{agile-boltztran} demonstrates that the solution of only one
transport equation for the neutrino distribution function can provide
accurate radiation transport solutions in the diffusion limit and
semi-transparent regime. Number and energy conservation are reasonably
well fulfilled and the description of hydrodynamics and radiation
transport add up to one consistent general relativistic finite difference
representation of radiation hydrodynamics in spherical symmetry \citep{Liebendoerfer_et_al_04}.
The approach is in principle extendible to two and three dimensions
and allows for adaptive zoning because it is based on the local description
of the transport equation in confined fluid elements. But consistency
is easily lost in higher dimensions \citep{Cardall_Mezzacappa_03}
and computer performance may become prohibitive in an implicit multidimensional
discrete ordinates approach.

\section{Conclusions}

\label{section_summary}We have compared two different approaches
to implement Boltzmann neutrino transport in spherically symmetric
radiation hydrodynamics simulations of stellar core collapse and postbounce
evolution. We performed calculations for two different progenitor
stars, the \( 13 \) M\( _{\odot } \) progenitor of \citet{Nomoto_Hashimoto_88}
and the \( 15 \) M\( _{\odot } \) progenitor of \citet{Woosley_Weaver_95}.
We present one Newtonian calculation (N13) with the minimum input
physics that leads to a plausible scenario after bounce and a second
relativistic calculation (G15) with the {}``standard'' physics used
in many recent supernova simulations. We find similar agreement in
both cases. The reduced complexity of the input physics in the N13
model helps to isolate differences in the implementation of the hydrodynamics
and the neutrino transport. We could improve the agreement by upgrading
the first order donor-cell advection scheme in the implicit hydrodynamics
code \textsc{agile} to a second order TVD advection scheme. The version
with first order advection led to a more pessimistic shock propagation
during the first \( 10 \) ms after bounce. It did, however, reveal
an interesting relationship between the transition of the propagating
discontinuity from a dynamical shock to an accretion front and the
almost coincident launch of the neutrino burst.

A neutrino burst radiated from an accretion front maintains a high
luminosity for a longer time than a neutrino burst produced by a dynamical
shock, because an accretion front compresses matter at steady-state
like conditions whereas the layer behind a dynamical shock gets diluted
quickly so that electron captures diminish on a short timescale. Therefore
less lepton number is lost in neutrinos from a dynamical shock which
rapidly crosses the neutrinospheres, but neutrinos extract more leptons
from the compressed matter behind the accretion front once the shock
has stalled (i.e., the postshock velocity has become negative). This
effect, however, turned out to produce transient differences only
for a few milliseconds in our simulations, and convergence of the
shock trajectories was found again after the shocks in both runs had
transformed to accretion fronts. While the optimistic N13 model with
only one neutrino flavor (\( \nu _{e} \) and \( \bar{\nu }_{e} \))
represents a case where the neutrinospheres are crossed by a dynamical
shock, the relativistic model G15 serves as an example where the shock
forms at a smaller enclosed mass due to the deeper general relativistic
potential and where additional losses occur by the emission of \( \mu  \)-
and \( \tau  \)-neutrinos from deeper layers. In this case the shock
turns into an accretion front before or at the time the neutrino burst
is launched.

The overall evolution of both models is in good agreement when simulations
with the two codes are compared. Differences in details were found,
e.g., a slightly different shock propagation in the early hydrodynamical
phase and more smearing of the composition interfaces in the outer
progenitor layers by artificial diffusion in the case of \textsc{agile}.
The luminosities in \textsc{vertex} tend to be slightly higher than
in \textsc{agile-boltztran} and the rms energies a little lower in
the N13 model. The approximation of general relativistic effects by
a modified gravitational potential in otherwise Newtonian hydrodynamics
in \textsc{vertex} is very accurate up to bounce. In comparison with
the general relativistic simulation of \textsc{agile-boltztran}, however,
a somewhat deeper potential associated with higher accretion rates
develops during the long-term postbounce evolution. The consequence
are larger neutrino luminosities and rms energies. But in general,
good qualitative and satisfactory quantitative agreement of all important
temporal and radial features was found also in the relativistic model.
Major differences can result from implementation-specific rather than
from method-specific details, e.g. from the former use of a low-order
advection scheme in \textsc{agile-boltztran} or from the specific
choice of the finite differencing in both codes.

We come to the conclusion that both methods work satisfactorily well
in this application and give comparable results. We determined similar
computational needs for our not thoroughly optimized codes. Standard
runs with \textsc{agile-boltztran} tend to consume slightly less computer
time. But standard runs with \textsc{vertex} have been performed with
better energy resolution and the angular resolution that can be achieved
at larger radii is out of reach for S\( _{N} \) methods. Hence, a
detailed comparison of CPU time requirements is not really meaningful.
Moreover, faster methods may have been developed in the meantime \citep{Burrows_et_al_00, Thompson_Burrows_Pinto_03}.
Rather than arguing about the {}``best'' method for a certain application,
we recommend to pursue a variety of feasible numerical approaches
for future astrophysical simulations, opening up the possibility of
independent mutual validation of the results. We hope that our comparison
provides a useful step towards quantitative modeling of a very complex
astrophysical problem.

\section*{Acknowledgments}

We thank R.~Buras and O. E. B. Messer for many important contributions
to {\sc vertex} and {\sc agile-boltztran} respectively, in particular
for implementing the three-flavor versions of the codes and coining
subroutines to calculate the neutrino pair and bremsstrahlung rates.
We are grateful for the most recent subroutine for ion-ion correlations
provided to us by N. Itoh and collaborators. We are greatly indebted
to Chris Fryer for suggesting and promoting this comparison. The Institute
for Nuclear Theory at the University of Washington is acknowledged
for its hospitality and the Department of Energy for support during
a visit of the Summer Program on Neutron Stars in 2001, during which
we started this work. AM and ML acknowledge funding by the NSF under
contract AST-9877130, the Oak Ridge National Laboratory, managed by
UT-Batelle, LLC, for the U.S. Department of Energy under contract
DE-AC05-00OR22725, and the DoE HENP SciDAC Program. HTJ and MR are
grateful for support by the Sonderforschungsbereich 375 on {}``Astroparticle
Physics'' of the Deutsche Forschungsgemeinschaft. The computations
of the Garching group were mostly performed on the CRAY T90 and CRAY
SV1ex of the John von Neumann Institute for Computing (NIC) in J\"{u}lich
and the computations of the Oak Ridge-Basel group on the CITA Itanium
I.


\begin{thebibliography}{Liebend\"orfer, Mezzacappa, \& Thielemann (2001)}
\bibitem[Bethe \& Wilson (1985)]{Bethe_Wilson_85}Bethe, H.~A. \& Wilson, J.~R. 1985, ApJ, 295, 14
\bibitem[Bludman \& Cernohorsky (1995)]{Bludman_Cernohorsky_95}Bludman, S.~A. \& Cernohorsky, J. 1995, Phys. Rep., 256, 37
\bibitem[Bruenn (1985)]{Bruenn_85}Bruenn, S.~W. 1985, ApJS, 58, 771
\bibitem[Bruenn \& Mezzacappa (1997)]{Bruenn_Mezzacappa_97}Bruenn, S.~W. \& Mezzacappa, A. 1997, Phys. Rev. D, 56, 7529
\bibitem[Bruenn, DeNisco, \& Mezzacappa (2001)]{Bruenn_DeNisco_Mezzacappa_01}Bruenn, S.~W., DeNisco, K.~R., \& Mezzacappa, A. 2001, ApJ, 560,
326
\bibitem[Buras et al. (2003)]{Buras_et_al_03}Buras, R., Rampp, M., Janka, H.-Th., \& Kifonidis, K. 2003, Phys.
Rev. Lett., 90, 241101
\bibitem[Burrows et al. (2000)]{Burrows_et_al_00}Burrows, A., Young, T., Pinto, P., Eastman, R., \& Thompson, T.~A.
2000, ApJ, 539, 865
\bibitem[Calder et al. (2002)]{Calder_et_al_02}Calder, A.~C., Fryxell, B., Plewa, T., Rosner, R., Dursi, L.~J.,
Weirs, V.~G., Dupont, T., Robey, H.~F., Kane, J.~O., Remington,
B.~A., Drake, R. P., Dimonte, G., Zingale, M., Timmes, F.~X., Olson,
K., Ricker, P., MacNeice, P., Tufo, H.~M. 2002, ApJS, 143, 201
\bibitem[Cardall \& Mezzacappa (2003)]{Cardall_Mezzacappa_03}Cardall,C.~Y., Mezzacappa, A. 2003, Phys. Rev. D, 68, 023006 
\bibitem[Cernohorsky (1994)]{Cernohorsky_94}Cernohorsky, J. 1994, ApJ, 433, 247
\bibitem[Colella \& Woodward (1984)]{Colella_Woodward_84}Colella, P. \& Woodward, P. 1984, J. Comp. Phys., 54, 174
\bibitem[Dorfi \& Drury (1987)]{Dorfi_Drury_87}Dorfi, E.~A., \& Drury, L.~O'C. 1987, J. Comput. Phys., 69, 175
\bibitem[Einfeldt (1988)]{Einfeldt_88}Einfeldt, B. 1988, SIAM Jour.~Numer.~Anal., 25, 294
\bibitem[Fryxell, M\"uller, \& Arnett (1989)]{Fryxell_Mueller_Arnett_89}Fryxell, B., M\"{u}ller, E., \& Arnett, W. 1989, Hydrodynamics and
Nuclear Burning, preprint MPA-449, Max Planck Institut f\"{u}r Astrophysik,
Garching
\bibitem[Horowitz (1997)]{Horowitz_97}Horowitz, C. 1997, Phys. Rev. D, 55, 4577
\bibitem[Itoh (1975)]{Itoh_75}Itoh, N. 1975, Prog. Theor. Phys., 54, 1580
\bibitem[Itoh et al. (2004)]{Itoh_et_al_04}Itoh, N., Asahara, R., Tomizawa, N., Wanajo, S., Nozawa, S. 2004,
astro-ph/0401488
\bibitem[Janka (2001)]{Janka_01}Janka, H.-Th. 2001, A\&A, 368, 527
\bibitem[Janka et al. (2004)]{Janka_et_al_04}Janka, H.-Th., Buras, R., Kifonidis, K., Plewa, T., \& Rampp, M. 2004,
in Stellar Collapse, ed. by Fryer, C.~L. (Dordrecht, the Netherlands:
Kluwer Academic Publishers), 65
\bibitem[Keil (1997)]{Keil_97}Keil, W. 1997, PhD thesis, Technische Universit\"{a}t M\"{u}nchen
\bibitem[Kifonidis (2000)]{Kifonidis_00}Kifonidis, K. 2000, PhD thesis, Technische Universit\"{a}t M\"{u}nchen
\bibitem[Lattimer \& Swesty (1991)]{Lattimer_Swesty_91}Lattimer, J. \& Swesty, F.~D. 1991, Nucl. Phys., A535, 331
\bibitem[Liebend\"orfer (2000)]{Liebendoerfer_00}Liebend\"{o}rfer, M. 2000, Ph.D. thesis (Basel: University of Basel)
\bibitem[Liebend\"orfer, Mezzacappa, \& Thielemann (2001)]{Liebendoerfer_Mezzacappa_Thielemann_01}Liebend\"{o}rfer, M., Mezzacappa, A., \& Thielemann, F.-K. 2001,
Phys. Rev., D63, 104003
\bibitem[Liebend\"orfer et al. (2001)]{Liebendoerfer_et_al_01}Liebend\"{o}rfer, M., Mezzacappa, A., Thielemann, F.-K., Messer,
O.~E.~B., Hix, W.~R., \& Bruenn, S.~W. 2001, Phys. Rev., D63,
103004
\bibitem[Liebend\"orfer, Rosswog, \& Thielemann (2002)]{Liebendoerfer_Rosswog_Thielemann_02}Liebend\"{o}rfer, M., Rosswog, S.~K., \& Thielemann, F.-K. 2002,
ApJS, 141, 229
\bibitem[Liebend\"orfer et al. (2004)]{Liebendoerfer_et_al_04}Liebend\"orfer, M., Messer, O.~E.~B., Mezzacappa, A., Bruenn, S.~W.,
Cardall, C. Y., \& Thielemann, F.-K. 2004, ApJS, 150, 263
\bibitem[Lindquist (1966)]{Lindquist_66}Lindquist, R.~W. 1966, Ann. Phys., 37, 487
\bibitem[May \& White (1966)]{May_White_66}May, M.~M., \& White, R.~H. 1966, Phys. Rev., 141, 1232
\bibitem[Messer (2000)]{Messer_00}Messer, O.~E.~B. 2000, Ph.D. thesis (Knoxville: University of Tennessee)
\bibitem[Mezzacappa \& Bruenn (1993a)]{Mezzacappa_Bruenn_93a}Mezzacappa, A. \& Bruenn, S.~W. 1993, ApJ 405, 669
\bibitem[Mezzacappa \& Bruenn (1993b)]{Mezzacappa_Bruenn_93b}Mezzacappa, A. \& Bruenn, S.~W. 1993, ApJ 405, 637
\bibitem[Mezzacappa \& Bruenn (1993c)]{Mezzacappa_Bruenn_93c}Mezzacappa, A. \& Bruenn, S.~W. 1993, ApJ 410, 740
\bibitem[Mezzacappa et al. (2001)]{Mezzacappa_et_al_01}Mezzacappa, A., Liebend\"{o}rfer, M., Messer, O.~E.~B., Hix, W.~R.,
Thielemann, F.-K., \& Bruenn, S.~W. 2001, PRL, 86, 1935
\bibitem[Mezzacappa \& Messer (1999)]{Mezzacappa_Messer_99}Mezzacappa, A. \& Messer, O.~E.~B. 1999, JCAM, 109, 281
\bibitem[Misner \& Sharp (1964)]{Misner_Sharp_64}Misner, C.~W., \& Sharp, D.~H. 1964, Phys. Rev., B136, 571
\bibitem[Myra et al (1987)]{Myra_et_al_87}Myra, E.~S., Bludman, S.~A., Hoffman, Y., Lichtenstadt, I., Sack,
N., \& Van Riper, K.~A. 1987, ApJ, 318, 744
\bibitem[Nomoto \& Hashimoto (1988)]{Nomoto_Hashimoto_88}Nomoto, K. \& Hashimoto, M. 1988, Phys. Rep., 163, 13
\bibitem[Plewa \& M\"uller (1999)]{Plewa_Mueller_99}Plewa, T. \& M\"{u}ller, E. 1999, A\&A, 342, 179
\bibitem[Plewa \& M\"uller (2001)]{Plewa_Mueller_01}Plewa, T. \& M\"{u}ller, E. 2001, Computer Physics Communications,
138, 101
\bibitem[Quirk (1994)]{Quirk_94}Quirk, J.~J. 1994, Int.~J.~Num.~Meth.~Fluids, 18, 555
\bibitem[Rampp (2000)]{Rampp_00}Rampp, M. 2000, PhD thesis, Technische Universit\"{a}t M\"{u}nchen
\bibitem[Rampp \& Janka (2000)]{Rampp_Janka_00}Rampp, M. \& Janka, H.~T. 2000, ApJL, 539, L33
\bibitem[Rampp \& Janka (2002)]{Rampp_Janka_02}Rampp, M. \& Janka, H.~T. 2002, A\&A, 396, 361
\bibitem[Smit, Cernohorsky, \& Dullemond (1997)]{Smit_Cernohorsky_Dullemond_97}Smit, J.~M., Cernohorsky, J., \& Dullemond, C.~P. 1997, A\&A, 325,
203
\bibitem[Sumiyoshi et al. (2000)]{Sumiyoshi_et_al_00}Sumiyoshi, K., Suzuki, H., Otsuki, K., Terasawa, M., \& Yamada, S.
2000, Publ. Astron. Soc. Jap., 52, 601
\bibitem[Takahashi, Witti, \& Janka (1994)]{Takahashi_Witti_Janka_94}Takahashi, K., Witti, J., \& Janka, H.-T. 1994, A\&A, 286, 857
\bibitem[Thompson, Burrows, \& Meyer (2001)]{Thompson_Burrows_Meyer_01}Thompson, T.~A., Burrows, A., \& Meyer, B.~S. 2001, ApJ, 562, 887
\bibitem[Thompson, Burrows, \& Pinto (2003)]{Thompson_Burrows_Pinto_03}Thompson, T.~A., Burrows, A., \& Pinto, P.~A. 2003, ApJ, 592, 434
\bibitem[Tscharnuter \& Winkler (1979)]{Tscharnuter_Winkler_79}Tscharnuter, W.~M., \& Winkler, K.~H. 1979, Comput. Phys. Comm.,
18, 171
\bibitem[Wanajo et al. (2001)]{Wanajo_et_al_01}Wanajo, S., Kajino, T., Mathews, G.~J., \& Otsuki, K. 2001, ApJ,
554, 578
\bibitem[Wilson (1985)]{Wilson_85}Wilson, J.~R. 1985, in Numerical Astrophysics, ed. by Centrella,
J.~M., LeBlanc, J.~M., \& Bowers, R.~L. (Boston: Jones and Bartlett),
422
\bibitem[Winkler, Norman, \& Mihalas (1984)]{Winkler_Norman_Mihalas_84}Winkler, K.-H., Norman, M.~L., \& Mihalas, D. 1984, J. Quant. Spectrosc.
Radiat. Transf., 31, 473
\bibitem[Woosley et al. (1994)]{Woosley_et_al_94}Woosley, S.~E., Wilson, J.~R., Mathews, G.~J., Hoffman, R.~D.,
\& Meyer, B.~S. 1994, ApJ, 433, 229
\bibitem[Woosley \& Weaver (1995)]{Woosley_Weaver_95}Woosley, S.~E., Weaver, T.~A. 1995, ApJS, 101, 181\end{thebibliography}
\end{document}